# Genetically designed biomolecular capping system for mesoporous silica nanoparticles enables receptor-mediated cell uptake and controlled drug release

Stefan Datz, Christian Argyo, Michael Gattner, Veronika Weiss, Korbinian Brunner, Johanna Bretzler, Constantin von Schirnding, Fabio Spada, Hanna Engelke, Milan Vrabel[§], Christoph Bräuchle, Thomas Carell, and Thomas Bein*

*Department of Chemistry, Nanosystems Initiative Munich (NIM), Center for Nano Science (CeNS), and Center for Integrated Protein Science Munich (CIPSM), University of Munich (LMU), Butenandtstr. 5-13, 81377 Munich, Germany. [§]Institute of Organic Chemistry and Biochemistry, Academy of Sciences of the Czech Republic, Email:bein@lmu.de.*

## Abstract

Effective and controlled drug delivery systems with on-demand release and targeting abilities have received enormous attention for biomedical applications. Here, we describe a novel enzyme-based cap system for mesoporous silica nanoparticles (MSNs) that is directly combined with a targeting ligand via bio-orthogonal click chemistry. The capping system is based on the pH-responsive binding of an aryl-sulfonamide-functionalized MSN and the enzyme carbonic anhydrase (CA). An unnatural amino acid (UAA) containing a norbornene moiety was genetically incorporated into CA. This UAA allowed for the site-specific bio-orthogonal attachment of even very sensitive targeting ligands such as folic acid and anandamide. This leads to specific receptor-mediated cell and stem cell uptake. We demonstrate the successful delivery and release of the chemotherapeutic agent Actinomycin D to KB cells. This novel nanocarrier concept provides a promising platform for the development of precisely controllable and highly modular theranostic systems.

## Introduction

The development of effective systems for targeted drug delivery combined with *on demand* release behavior can be considered one of the grand challenges in nanoscience. In particular, porous nanocarriers with high drug loading capacity, immunological stealth behavior and tunable surface properties are promising candidates for biomedical applications such as cancer therapy and bioimaging.[1-4] Specifically, multifunctional mesoporous silica nanoparticles (MSNs) have great potential in drug delivery applications due to their attractive porosity parameters and the possibility to conjugate release mechanisms for diverse cargos[5, 6] including gold nanoparticles,[7, 8] iron oxide nanocrystals,[9] bio-macromolecules,[10, 11] enzymes,[12] and polymers.[13] Control over a stimuli-responsive cargo release can be achieved via different trigger mechanisms such as redox reactions,[14] pH changes,[15] light-activation,[5] or change in temperature.[6] Drug delivery vehicles equipped with acid-sensitive capping mechanisms are highly desirable for acidified target environments such as the translation from early to late endosomes, tumors, or inflammatory tissues.

Here, we present genetically designed enzyme-capped MSNs that combine two important prerequisites for advances in drug delivery, namely stimuli-responsive drug release and specific cell targeting (Scheme 1). Specifically, these pH-responsive MSNs consist of a capping structure based on carbonic anhydrase (CA). CA is a model enzyme abundant in humans and animals and generally catalyzes the hydration of carbon dioxide and the dehydration of bicarbonate.[16] It is attached to the silica nanoparticle surface via aryl sulfonamide groups. As its natural inhibitor sulfonamide groups strongly bind to the active site of the CA. This enzyme-sulfonamide binding is reversible depending on the pH, where an acidic medium causes protonation of the sulfonamide, resulting in cleavage of the coordination bond and access to the porous system.[17] The CA gatekeepers were used to exploit the endosomal pH change as an internal cellular trigger and to gain control over the release of cargo molecules from the mesoporous system.

This stimuli-responsive capping system on MSNs was combined with cell targeting specificity via a bio-orthogonal click chemistry approach. Targeting ligands provide specific binding to certain cell membrane receptors allowing for an enhanced and distinctive cellular uptake of such modified nanocarriers. For example, various cell receptors are overexpressed on cancer cells, which can lead to a preferential receptor-mediated endocytosis of modified MSNs. For the attachment of such targeting ligands exclusively to the outer periphery of the enzyme gatekeepers, we exploited a recently developed method that takes advantage of the Pyrrolysine *amber suppression* system followed by bio-orthogonal copper-free click chemistry.[18-20] This system has already been utilized in applications such as optical gene control.[21] To the best of our knowledge, this is the first time the Pyrrolysine *amber suppression* system is used in a combination with porous nanocarriers for specific cell recognition and drug delivery. The incorporation of an unnatural amino acid (UAA) containing a norbornene moiety into CA provides a bio-orthogonal reaction pathway by covalently attaching tetrazine-modified targeting ligands.[22, 23] It has recently been demonstrated that norbornene-tetrazine click chemistry is a favorable synthesis strategy over various other methods including thiol-maleimide reaction and amide formation due to extremely mild and biocompatible reaction conditions and higher selectivity.[24] Here, copper-free click chemistry of norbornene-modified human carbonic anhydrase II with targeting ligands was performed to prepare folate- and anandamide-modified multifunctional mesoporous silica nanocarriers.[25] The anandamide is, due the cis-configured double bonds, a particularly sensitive receptor ligand that requires extremely mild coupling conditions. The targeting system based on folate-modified silica nanocarriers was studied on KB cancer cells, which are known to overexpress the folate receptor FR-α.[5, 26] The targeting system based on anandamide-modified particles was tested on neural stem cells. The combination of on-demand release and specific receptor-mediated cell uptake properties within one multifunctional mesoporous silica nanocarrier system, containing biomolecular valves based on carbonic anhydrase, is anticipated to offer promising potential for controlled drug delivery applications including cancer therapy.

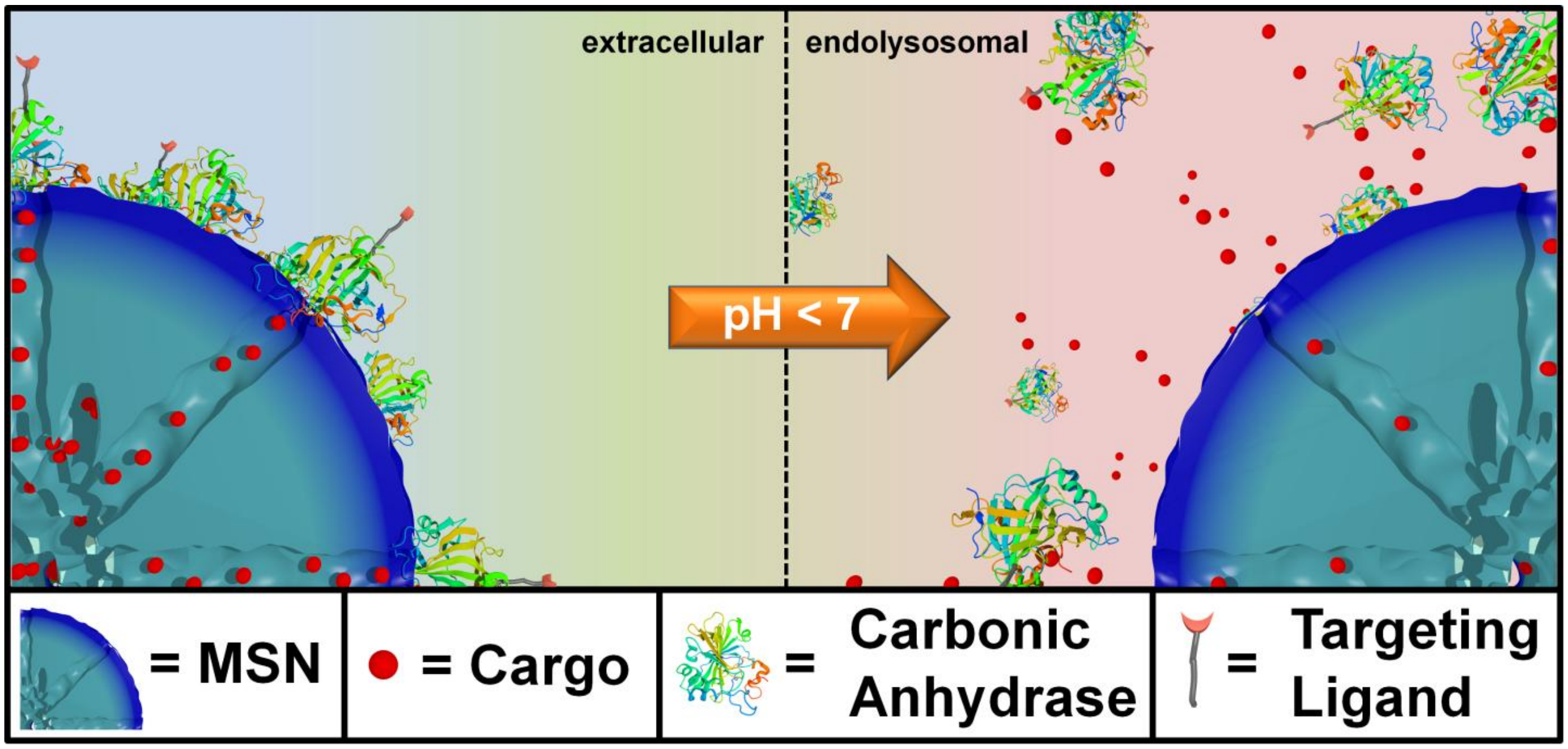

**Scheme 1: Schematic illustration of the genetically designed biomolecular pore gating system providing a pH-responsive drug release from mesoporous silica nanoparticles (MSNs). Aryl sulfonamide functionalized MSNs offer pH-dependent reversible attachment of the bulky enzyme carbonic anhydrase, which efficiently blocks the pore entrances to prevent premature cargo release. Furthermore, specific cancer cell targeting can be achieved via site-specific modification of a genetically incorporated norbornene amino acid in the biomolecular gatekeepers.**

## Results and Discussion

pH-Responsive MSNs with an average particle size of 150 nm (average pore diameter: 3.8 nm) containing biomolecular valves based on the enzyme carbonic anhydrase (CA, hydrodynamic diameter: 5.5 nm) were synthesized via a delayed co-condensation approach.[27] An outer functional shell consisting of benzene sulfonamide groups (pHSA) acts as an anchor point for the enzymatic gatekeepers. The formation of the inhibitor-enzyme complex (phSA-CA) leads to a dense coating at the external particle surface (MSN-phSA-CA). The characterization of the particle system by dynamic light scattering (DLS), zeta potential measurements, transmission electron microscopy (TEM), nitrogen sorption isotherms, infrared and Raman spectroscopy confirms the successful synthesis of carbonic anhydrase-coated MSNs. *In vial* release experiments demonstrate efficient sealing of the pores with carbonic anhydrase acting as a bulky gatekeeper, preventing premature cargo release and allowing for release upon acid-induced detachment of the capping system (for detailed information about synthesis and characterization see SI).

For efficient receptor-mediated cancer cell uptake and selective drug delivery a targeting ligand needs to be implemented. Since the particle surface is covered with bulky enzymes (CA), we aimed for the attachment of the targeting moieties directly to the outer periphery of the enzyme, in order to be accessible for cell receptors. For this approach to be successful, the site of targeting ligand attachment on the enzyme is of key importance. Ideally it should be positioned opposite of the binding site of the enzyme, to prevent blocking of the active site and thus leakage of the capping system. However, site-specific chemical modifications of proteins are highly challenging. Several methods, such as the reaction of thiol groups with maleimide or of lysine chains with activated esters, lack specificity. A more specific method is the incorporation of unnatural amino acids into the protein.[18, 28, 29] Among others, the genetic incorporation of UAAs bearing side chains with alkyne,[30, 31] trans-cyclooctene,[32] cyclooctyne[33] or norbornene[18, 22] functionalities has been reported previously. Subsequently these residues can be modified specifically and bio-orthogonally, for example by reverse electron-demanding Diels-Alder reactions with tetrazines.[23, 24, 32] The natural PylRS/$tRNA_{Pyl}$ pair is perfectly suitable to genetically incorporate UAAs due to its orthogonality to common expression strains. Recently, a norbornene-containing Pyl analogue (Knorb) has been developed by some of us.[18, 19] Here, the synthesis of norbornene-functionalized human carbonic anhydrase II (HCA) was accomplished similar to a previously described procedure yielding HCA H36Knorb.[34] The correct position of the UAA was confirmed by tryptic digestion of the protein followed by HPLC-MS/MS analysis (see SI). HCA H36Knorb carrying norbornene on the opposite face of its phSA-binding site was bound to phSA-MSN and then treated with an excess of folate-$PEG_{2000}$-tetrazine (Figure 1) or anandamide-tetrazine. The excess of the tetrazine reagent could be easily removed by centrifugation of the nanoparticles followed by washing. The efficiency of the folate-targeting system was examined on KB-cells presenting either free or blocked FA-receptors (Figure 1). For visualization, the cell membrane of the KB cells was stained with WGA488 (green), and the particles were labeled with Atto633 (red). In Figure 1c-e we present the folic acid receptor blocked cells that were incubated with particles between 2 and 8 h. With increasing incubation time, only a few particles were internalized and unspecific cell uptake was observed only to a minor degree. In contrast, the cells with available folic acid receptor on their surface (Figure 1f-h) exhibit a significant and increasing uptake behavior and a considerably higher degree of internalized particles. Thus we could confirm the successful application of bioorthogonal modification of a capping enzyme to act as targeting ligand. We also proved, that the here described genetically modified enzyme capping strategy can be used to attach even sensitive ligands like arachidonic acid via mild click-chemistry conditions e.g. for the site-specific targeting of neural stem cells.[35] We tested the anandamide-targeting system on neural stem cells. Neural stem cells have anandamide receptors and successfully internalized the anandamide-particles (see SI).

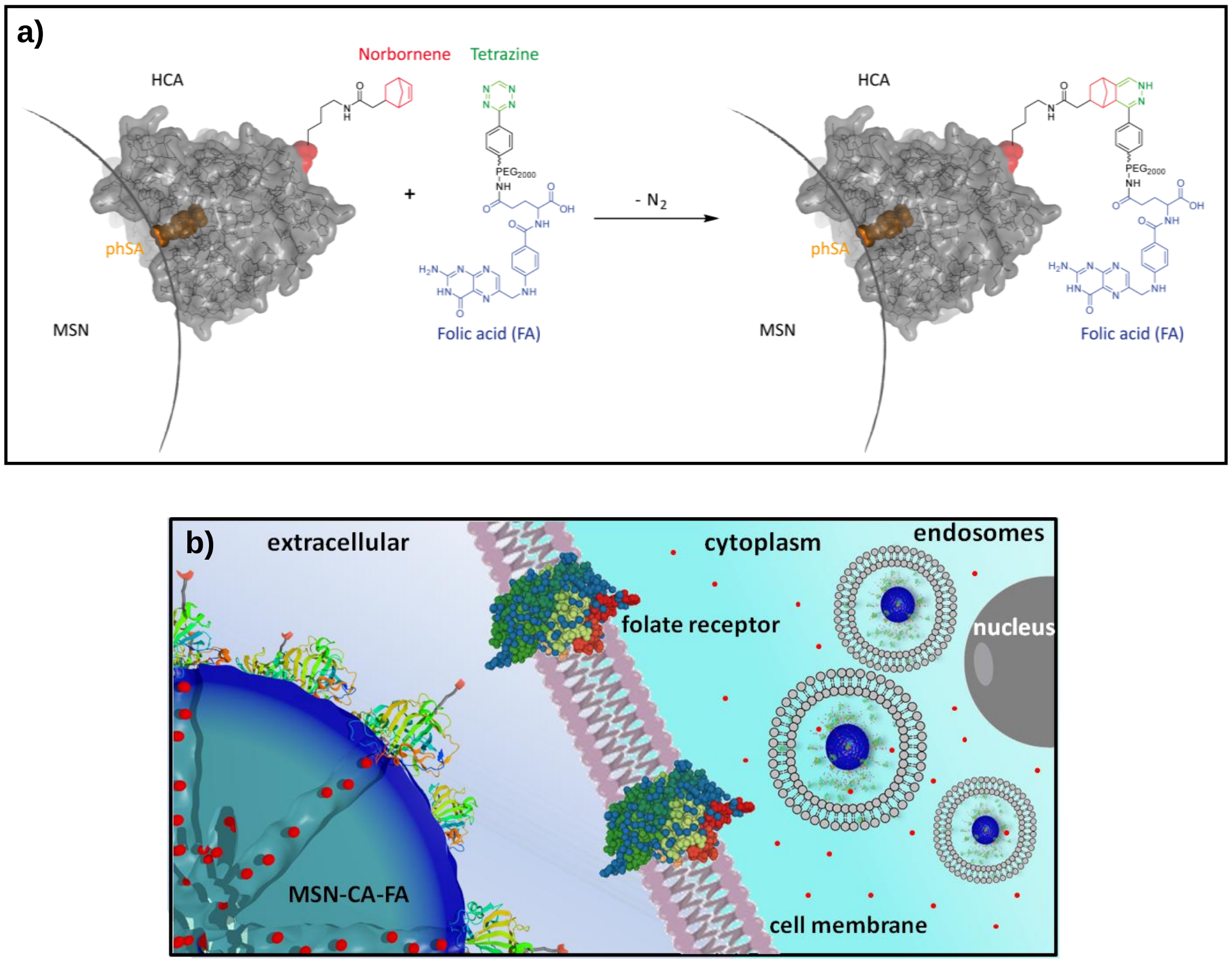

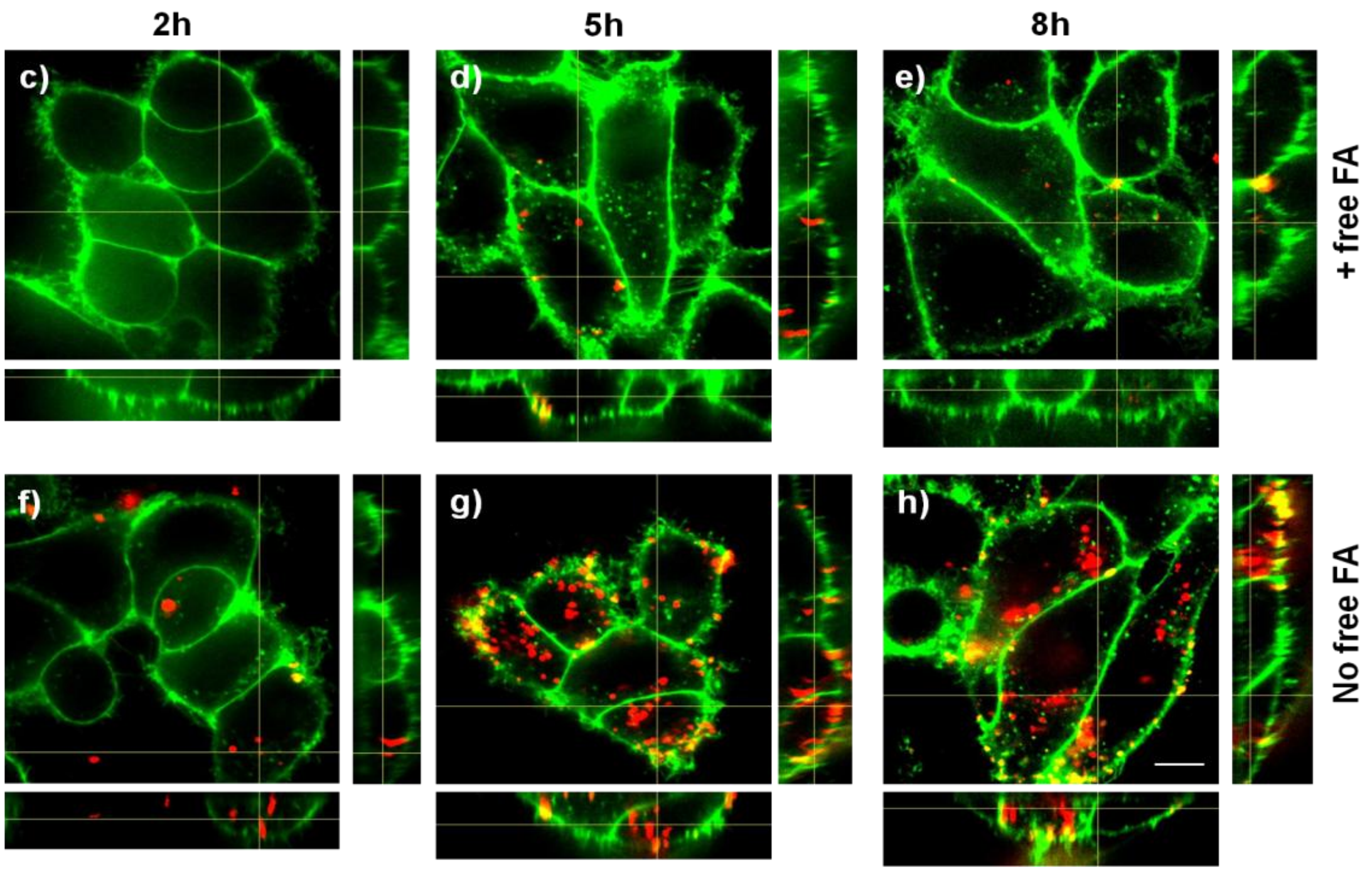

**Figure 1: a) Norbornene-functionalized carbonic anhydrase (HCA H36Norb) with indicated functionalization site (red) and active site (blue) is able to react in a reversed-electron-demand Diels-Alder reaction with a folate-PEG2000-tetrazine derivative to give HCA-FA. b) Schematic receptor-mediated uptake of folate-functionalized MSN-CA nanoparticles. c-e) Nonspecific and f-h) receptor-mediated endocytosis of MSN-phSA-CA-FA (red) by KB cells (WGA488 membrane staining, green). A specific receptor-mediated cell uptake was observed for MSN-phSA-CA-FA with KB cells (not pre-incubated with FA) after 5 and 8 h incubation at 37 °C (g/h). Incubation of MSN-phSA-CA-FA with FA-pre-incubated KB cells for 2, 5, 8 h at 37 °C showed only minor unspecific cellular uptake over all incubation times (c-e). The scale bar represents 10 µm.**

*In vitro* release experiments with the model cargo DAPI demonstrate a substantial time-dependent release behavior from the mesopores of our nanocarrier system in HeLa cells (Figure S10). Additional co-localization experiments showed the localization of CA-capped nanoparticles in acidic cell compartments after endocytosis (Figure S11). To examine the ability of our newly developed MSN drug delivery system to transport chemotherapeutics and to affect cells with their cargo, we incorporated Actinomycin D (AmD), a cytostatic antibiotic, dissolved in DMSO. Free AmD is membrane permeable and induced an uncontrolled cell death within a few hours. MSN-phSA-CA provided intracellular AmD release and caused efficient cell death after 24 h. The delayed reaction demonstrates that AmD was delivered in a controlled manner via the particles and released only after acidification of the endosome and subsequent de-capping of the gate-keeper CA. In Figure 2 cell death is visualized by a caspase 3/7 stain - a marker for apoptotic/dead cells. Control particles loaded with pure DMSO did not induce cell death at all, nor did the supernatant solution after particle separation via centrifugation (Figure 2i-l). This experiment shows the great potential of the MSN-phSA-CA system to efficiently deliver chemotherapeutics to cancer cells. The pH-responsive genetically modified capping system provides the ability to act as a general platform for different targeting ligands and cargos.

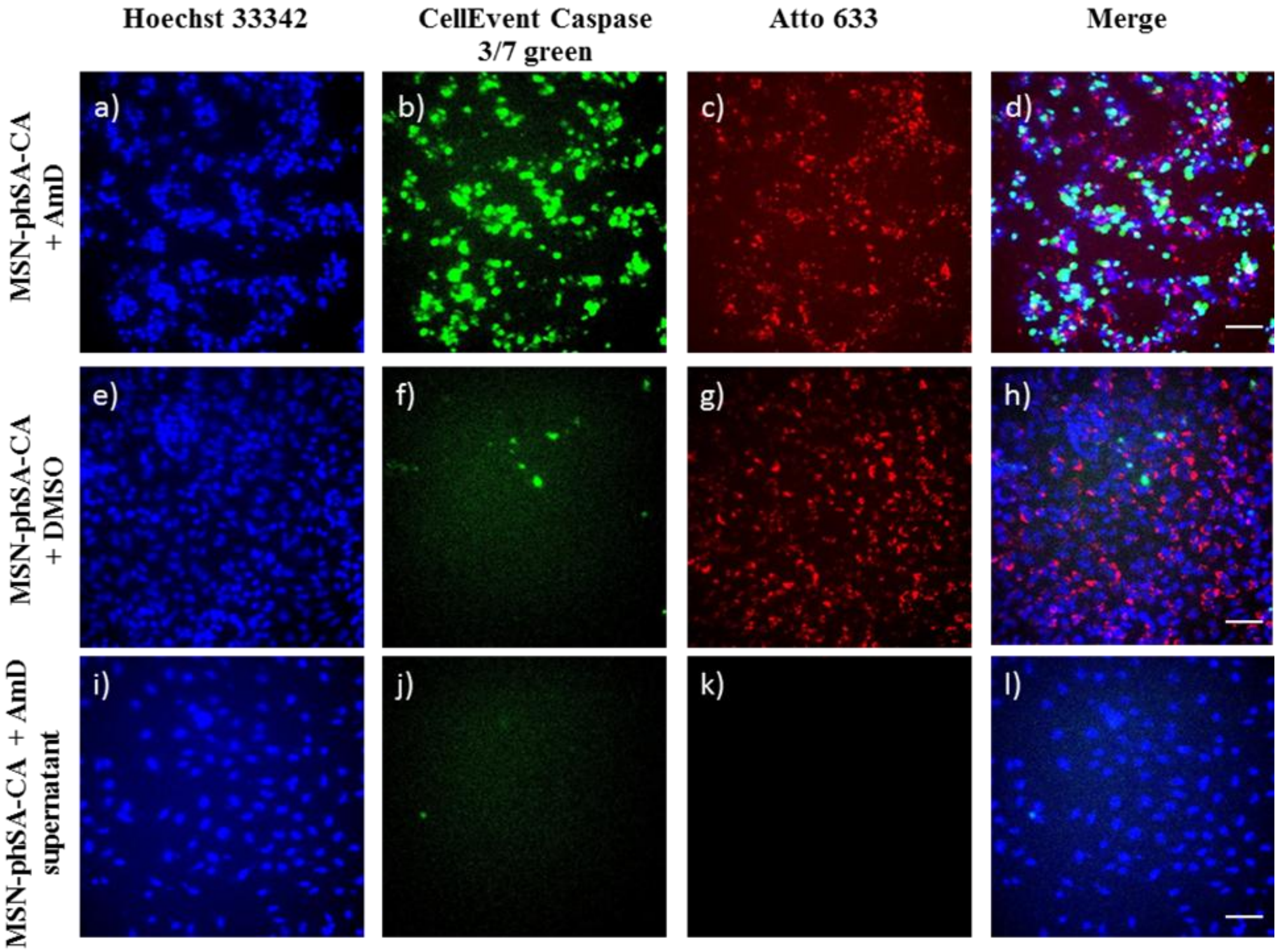

**Figure 2: Representative fluorescence microscopy images of HeLa cells incubated with MSN-phSA-CA nanoparticles loaded with Actinomycin D (AmD; a-d) or DMSO (e-h) and labeled with Atto 633 (red) after 24 h of incubation. As a control, the supernatant of AmD loaded particles after particle separation was incubated with the cells (i-l). Cell nuclei were stained with**

**Hoechst 33342 (blue). For live/dead discrimination CellEvent caspase 3/7 (green) was used. Due to activation of caspase-3/7 in apoptotic cells, DNA can be stained after cleavage of the DNA-binding dye from a binding-inhibiting peptide. MSNs were efficiently taken up by cells (c/d and g/h). Cell death can only be observed for cells treated with AmD loaded MSN-phSA-CA after 24 h of incubation (increased DNA staining in green) (b). In contrast, nanoparticles loaded with DMSO or the sample supernatant do not induce significant apoptosis (almost no DNA-staining) (f and j). The scale bars represent 50 µm.**

## Conclusions

We conclude that the novel capping system concept based on pH-responsive detachment of carbonic anhydrase combined with folic acid as targeting ligand allows for highly controllable drug release from porous nanocarriers. Our drug delivery system provides an on-demand release mechanism shown by *in vial* and *in vitro* cargo release experiments. The multifunctional MSNs were efficiently endocytosed in cancer cells and could be located in acidic cell compartments where they released their cargo. Furthermore, the system has an on-board targeting mechanism as demonstrated in additional *in vitro* experiments. The targeting mechanism is attached at a specific site of the capping enzyme preventing interference with the closure mechanism. Our newly developed pH-responsive gatekeepers with genetically designed targeting functions provide a promising platform for the design of versatile and modular drug delivery systems.

## Acknowledgements

The authors acknowledge financial support from the Deutsche Forschungsgemeinschaft (DFG) in the context of SFB 749 and SFB 1032, the Excellence Clusters Nanosystems Initiative Munich (NIM) and Center for Integrated Protein Science Munich (CIPSM), and the Center for Nano Science (CeNS). We thank Dr. Bastian Rühle for 3D graphics design.

## Supporting Information

Experimental part, further characterization of functionalized MSNs, results of additional cell studies, cloning, expression and purification of HCA H36Knorb.

# Supporting Information

## Experimental Part

**Materials.** Tetraethyl orthosilicate (TEOS, Fluka, > 99 %), triethanolamine (TEA, Aldrich, 98 %), cetyltrimethylammonium chloride (CTAC, Fluka, 25 % in $H_2O$), (3-mercaptopropyl)-triethoxysilane (MPTES, Sigma Aldrich, > 80 %), 6-maleimidohexanoic acid *N*-hydroxysuccinimide ester (Fluka, > 98 %), bovine carbonic anhydrase (bCA, Sigma, > 95 %), 4-(2-aminoethyl)benzenesulfonic acid (Aldrich, 98 %), folic acid (FA, Sigma Aldrich, ≥ 97 %,), 4,6-Diamidino-2-phenylindole dihydrochloride (DAPI, Sigma-Aldrich, ≥ 98 %), CellEvent™ Caspase-3/7 Green Detection Reagent (lifeTechnologies), Hoechst 33342, Trihydrochloride, Trihydrate (lifeTechnologies), Wheat Germ Agglutinin, Alexa Fluor® 488 Conjugate (lifeTechnologies), CellLight© Early Endosome-GFP, Late Endosome-GFP, and Lysosome-GFP, BacMam 2.0 (lifeTechnologies), Atto 633 maleimide (ATTO-TEC), ammonium nitrate ($NH_4NO_3$, Aldrich), ammonium fluoride ($NH_4F$, Aldrich), hydrochloric acid (37 %), fluorescein disodium salt dihydrate (Aldrich, 90 %), and Hank´s balanced salt solution (HBSS-buffer, Sigma Aldrich) were used as received. Ethanol (EtOH, absolute, Aldrich), DMSO and dimethylformamide (DMF, dry, Aldrich) were used as solvent without further purification. Bidistilled water was obtained from a millipore system (Milli-Q Academic A10). Citric-acid phosphate buffer (CAP-buffer, pH 5.5) was freshly prepared by carefully mixing a certain amount of disodium hydrogen phosphate ($Na_2HPO_4$, 0.2 M in $H_2O$) and citric acid (0.2 M in $H_2O$) to adjust a pH value of 5.5. Subsequently, the solution was diluted with bidistilled $H_2O$ to a total volume of 500 mL.

**Synthesis of thiol-functionalized MSNs (MSN-SH).** A mixture of TEOS (1.92 g, 9.22 mol) and TEA (14.3 g, 95.6 mmol) was heated to 90 °C for 20 min under static conditions in a polypropylene reactor. Then, a preheated (60 °C) mixture of CTAC (2.41 mL, 1.83 mmol, 25 % in $H_2O$) and $NH_4F$ (100 mg, 0.37 mmol) in bidistilled $H_2O$ (21.7 g, 1.21 mol) was added and the resulting reaction mixture was stirred vigorously (700 rpm) for 30 min while cooling down to room temperature. Afterwards, TEOS (18.2 mg, 92 µmol) and MPTES (18.1 mg, 92 µmol) were premixed briefly before addition to the reaction mixture. The final reaction mixture was stirred over night at room temperature. After dilution with absolute ethanol (100 mL), the nanoparticles were collected by centrifugation (19,000 rpm, 43,146 rcf, 20 min) and redispersed in absolute ethanol. Template extraction was performed in an ethanolic solution of MSNs (100 mL) containing $NH_4NO_3$ (2 g) which was heated at reflux conditions (90 °C oil bath) for 45 min. This was followed by a second extraction step (90 mL absolute ethanol and 10 mL hydrochloric acid (37 %)) under reflux conditions for 45 min (the material was washed with absolute ethanol after each extraction step and collected by centrifugation); finally the particles were redispersed in absolute ethanol and stored as colloidal suspension.

**Synthesis of sulfonamide-functionalized MSNs (MSN-phSA).** For the covalent attachment of a sulfonamide derivative to the external particle surface, a thiol-reactive linker was synthesized. 6-maleimidohexanoic acid *N*-hydroxysuccinimide ester (mal-C6-NHS, 10 mg, 33 µmol) was dissolved in DMF (500 µL, dry) and was added to an ethanolic solution (15 mL) containing 4-(2-aminoethyl)benzene sulfonamide (6.7 mg, 33 µmol). The resulting reaction mixture was stirred for 1 h at room temperature. Afterwards, thiol-functionalized silica nanoparticles (MSN-SH, 100 mg) in absolute ethanol (10 mL) were added and the mixture was stirred over night at room temperature. Subsequently, the particles were collected by centrifugation (19,000 rpm, 41,146 rcf, 20 min), washed twice with absolute ethanol and were finally redispersed in ethanol (15 mL) to obtain a colloidal suspension.

**Cargo loading and particle capping.** MSNs (MSN-phSA, 1 mg) were immersed in an aqueous solution of fluorescein (1 mL, 1 mM), DAPI (500 µL, 14.3 mM) or Actinomycin D (500 µL [14 v% DMSO], 140 µM) and stirred over night or for 1 h, respectively. After collection by centrifugation (14,000 rpm, 16,837 rcf, 4 min), the loaded particles were redispersed in a HBSS buffer solution (1 mL) containing carbonic anhydrase (1 mg) and the resulting mixture was allowed to react for 1 h at room temperature under static conditions. The particles were thoroughly washed with HBSS buffer (4 times), collected by centrifugation (5,000 rpm, 2,200 rcf, 4 min, 15 °C), and finally redispersed in HBSS buffered solution.

**Synthesis of Knorb.** The norbornene containing amino acid Knorb was synthezised as described in Ref.[1]

**Mutagenesis of pACA_HCA H36amber.** Adapted from Ref. [2] with permission from The Royal Society of Chemistry. The amber codon (TAG) was introduced into the expression vector pACA_HCA[3] at position His36 of the human carbonic anhydrase II gene by blunt end site directed mutagenesis using the primers *forward HCA H36amber* and *reverse HCA H36amber* (see Table S1).

**Table S1**: Sequences of the used primers for the generation of expression vector pACA_HCA H36amber. The introduced Amber codon is shown in bold.

| **Name** | **Sequence** |
|---|---|
| forward HCA H36amber | 5’phosph GTT GAC ATC GAC ACT **TAG** ACA GCC AAG TAT GAC |
| reverse HCA H36amber | 5’phosph AGG GGA CTG GCG CTC TCC CTT GG |

**Expression of norbornene-containing HCA.** Adapted from Ref.[2] with permission from The Royal Society of Chemistry. The expression vector pACA_HCA G131amber was transformed together with pACyc_pylRS Norb, 3xpylT[1] which contains the genes of the triple mutant of PylRS and three copies of *pyl*T in *E. coli* BL21(DE3) cells (NEB). 1 L of LB medium containing 34 mg/L chloramphenicol, 100 mg/L carbenicillin and 2 mM norbornene amino acid Knorb was inoculated with 10 mL of an overnight culture. The cells were stirred at 37 °C until an $OD_{600}$ of 0.9. At this optical density 1 mM $ZnSO_4$ and 0.1 mM IPTG were added to induce the expression of the HCA H36amber gene. After further 10 h at 37 °C the cells were harvested and stored at -20 °C until further use. The harvested cells were resuspended in washing buffer (25 mM Tris; 50 mM $Na_2SO_4$; 50 mM $NaClO_4$; pH 8.8) and disrupted by French Press procedure. The supernatant of the centrifuged lysate was used for sulfonamide affinity protein purification using an ÄKTA purifier system. The self-packed 3 mL column of *p*-Aminomethylbenzenesulfonamide-Agarose resin (Sigma-Aldrich, A0796) was equilibrated with washing buffer. After binding (0.75 mL/min) of the protein solution, the column was washed with 7 column volumes of washing buffer. HCA was eluted by lowering the pH by elution buffer (100 mM NaOAc; 200 mM $NaClO_4$; pH 5.6). The protein containing fractions were combined, analyzed by SDS-PAGE, dialyzed against water and lyophilized. Typical yields of the pure norbornene amino acid Knorb containing protein HCA H36Knorb were 20 mg/L expression medium.

**Tryptic digestion and MS/MS of norbornene-containing HCA.** Adapted from Ref. [2] with permission from The Royal Society of Chemistry. The sequence of HCA II is shown in Table S2. Position His36 which was chosen for the incorporation of amino acid Knorb is shown in red. The peptide generated after tryptic digestion is emphasized in bold letters. Figure S1 shows the corresponding MS/MS spectrum. Table S3 shows the expected and identified MS/MS fragments of the relevant tryptic peptide.

**Table S2**: Amino acid sequence of HCA II.

| 10 | 20 | 30 | 40 | 50 | 60 |
|---|---|---|---|---|---|
| MAHHWGYGKH | NGPEHWHKDF | PIAKGER**QSP** | **VDIDTHTAK**Y | DPSLKPLSVS | YDQATSLRIL |
| 70 | 80 | 90 | 100 | 110 | 120 |
| NNGHAFNVEF | DDSQDKAVLK | GGPLDGTYRL | IQFHFHWGSL | DGQGSEHTVD | KKKYAAELHL |
| 130 | 140 | 150 | 160 | 170 | 180 |
| VHWNTKYGDF | CKAVQQPDGL | AVLGIFLKVG | SAKPGLQKVV | DVLDSIKTKG | KSADFTNFDP |
| 190 | 200 | 210 | 220 | 230 | 240 |
| RGLLPESLDY | WTYPGSLTTP | PLLESVTWIV | LKEPISVSSE | QVLKFRKLNF | NGEGEPEELM |
| 250 | 260 | | | | |
| VDNWRPAQPL | KNRQIKASFK | | | | |

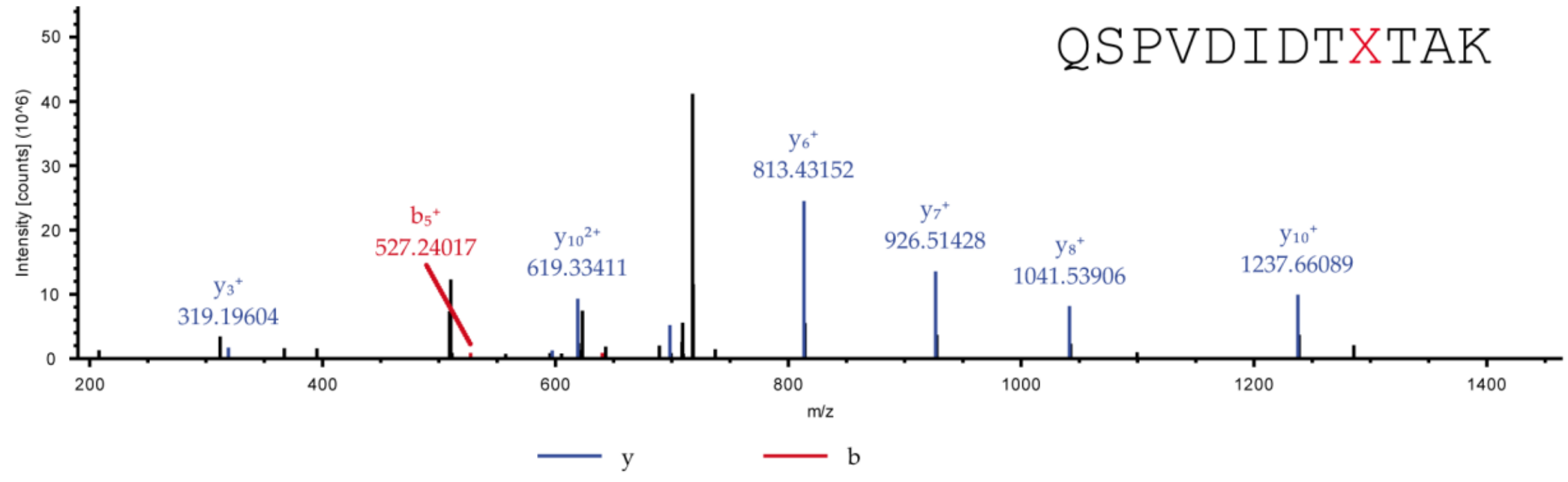

**Figure S1**: MS/MS spectrum of the tryptic peptide QSPVDIDTXTAK (X = **4**). Parent ion: $[M+2H]^{2+}_{calc.}$ = 726.8829, $[M+2H]^{2+}_{obs.}$ = 726.8807 (ΔM = 3 ppm).

**Table S3**: Expected and identified MS/MS fragments of the tryptic peptide QSPVDIDTXTAK (X = Knorb). Identified fragments are shown in red for b ions and blue for y ions.

| #1 | $b^+$ | Seq. | $y^+$ | #2 |
|---|---|---|---|---|
| 1 | 129.06586 | Q | | 12 |
| 2 | 216.09789 | S | 1324.69949 | 11 |
| 3 | 313.15066 | P | 1237.66746 | 10 |
| 4 | 412.21908 | V | 1140.61469 | 9 |
| 5 | 527.24603 | D | 1041.54627 | 8 |
| 6 | 640.33010 | I | 926.51932 | 7 |
| 7 | 755.35705 | D | 813.43525 | 6 |
| 8 | 856.40473 | T | 698.40830 | 5 |
| 9 | 1134.56774 | X | 597.36062 | 4 |
| 10 | 1235.61542 | T | 319.19761 | 3 |
| 11 | 1306.65254 | A | 218.14993 | 2 |
| 12 | | K | 147.11281 | 1 |

**Click chemistry of norbornene-containing hCA.** MSNs (MSN-phSA, 0.5 mg) were immersed in 500 µL HBSS buffer solution and 0.5 mg norbornene-containing hCA were added. In the meantime, 2.5 µg tetrazin *p*-benzylamine (DMSO stock solution, 0.92 mg/mL) and 0.41 mg NHS-PEG$_{2000}$-FA were mixed in 100 µL HBSS and stirred overnight in the dark at room temperature. The solutions were mixed afterwards and stirred for two hours, washed several times and redispersed in 1 mL HBSS buffer. Subsequently, 1 µL Atto633mal (DMF stock solution, 0.5 mg/mL) was added and the mixture was stirred for 1 hour. The particles were thoroughly washed with HBSS buffer (4 times), collected by centrifugation (5,000 rpm, 2,200 rcf, 4 min, 15 °C), and finally redispersed in HBSS buffered solution.

**Characterization.** DLS and zeta potential measurements were performed on a Malvern Zetasizer Nano instrument equipped with a 4 mW He-Ne-Laser (633 nm) and an avalanche photodiode detector. DLS measurements were directly recorded in diluted colloidal suspensions of the particles at a concentration of 1 mg/mL. Zeta potential measurements were performed using the add-on Zetasizer titration system (MPT-2) based on diluted NaOH and HCl as titrants. For this purpose, 1 mg of the particles was diluted in 10 mL bi-distilled water. Thermogravimetric analyses (TGA) of the bulk extracted samples (approximately 10 mg) were recorded on a Netzsch STA 440 C TG/DSC. The measurements proceeded at a heating rate of 10 °C/min up to 900 °C in a stream of synthetic air of about 25 mL/min. Nitrogen sorption measurements were performed on a Quantachrome Instrument NOVA 4000e at -196 °C. Sample outgassing was performed for 12 hours at a vacuum of 10 mTorr at RT. Pore size and pore volume were calculated by a NLDFT equilibrium model of $N_2$ on silica, based on the desorption branch of the isotherms. In order to remove the contribution of the interparticle textural porosity, pore volumes were calculated only up to a pore size of 8 nm. A BET model was applied in the range of 0.05 – 0.20 $p/p_0$ to evaluate the specific surface area. Infrared spectra of dried sample powder were recorded on a ThermoScientific Nicolet iN10 IR-microscope in reflexion-absorption mode with a liquid-$N_2$ cooled MCT-A detector. Raman spectroscopy measurements were performed on a confocal LabRAM HR UV/VIS (HORIBA Jobin Yvon) Raman microscope (Olympus BX 41) with a SYMPHONY CCD detection system. Measurements were performed with a laser power of 10 mW at a wavelength of 633 nm (HeNe laser). Dried sample powder was directly measured on a coverslip. UV/VIS measurements were performed on a Perkin Elmer Lambda 1050 spectrophotometer equipped with a deuterium arc lamp (UV region) and a tungsten filament (visible range). The detector was an InGaAs integrating sphere. Fluorescence spectra were recorded on a PTI spectrofluorometer equipped with a xenon short arc lamp (UXL-75XE USHIO) and a photomultiplier detection system (model 810/814). The measurements were performed in HBSS buffer solution at 37 °C to simulate human body temperature. For time-based release experiments of fluorescein a custom made container consisting of a Teflon tube, a dialysis membrane (ROTH Visking type 8/32, MWCO 14,000 g/mol) and a fluorescence cuvette was used. The excitation wavelength was set to $\lambda = 495$ nm for fluorescein-loaded MSNs. Emission scans (505 – 650 nm) were performed every 5 min. All slits were adjusted to 1.0 mm, bandwidth 8 nm). Mass spectra were recorded a *Thermo LTQ-Orbitrap XL*. For analytical HPLC separations of protein and peptide samples with subsequent MS a *Dionex Ultimate 3000 Nano* HPLC was used. Acetonitrile of LC-MS grade was purchased from *Carl Roth GmbH + Co. KG*. Water was purified by a Milli-Q Plus system from *Merck Millipore*.

**Synthesis of Anandamide-tetrazine**

Chemicals were purchased from *Sigma-Aldrich*, *Fluka* or *Acros* and used without further purification. Solutions were concentrated *in vacuo* on a *Heidolph* rotary evaporator. The solvents were of reagent grade and purified by distillation. Chromatographic purification of products was accomplished using flash column chromatography on *Merck* Geduran Si 60 (40-63 μM) silica gel (normal phase). Thin layer chromatography (TLC) was performed on *Merck* 60 (silica gel F254) plates. Visualization of the developed chromatogram was performed using fluorescence quenching or staining solutions. $^{1}H$ and $^{13}C$ NMR spectra were recorded in deuterated solvents on *Bruker ARX 300*, *Varian VXR400S*, *Varian Inova 400* and *Bruker AMX 600* spectrometers and calibrated to the residual solvent peak. Multiplicities are abbreviated as follows: s = singlet, d = doublet, t = triplet, q = quartet, m = multiplet, br. = broad. ESI spectra and high-resolution ESI spectra were obtained on the mass spectrometers *Thermo Finnigan* LTQ FT-ICR. IR measurements were performed on *Perkin Elmer Spectrum BX FT-IR* spectrometer (Perkin Elmer) with a diamond-ATR (Attenuated Total Reflection) setup. Repetencys are given in cm-1. The intensities are abbreviated as follows: vs (very strong), s (strong), m (medium), w (weak), vw (very weak).

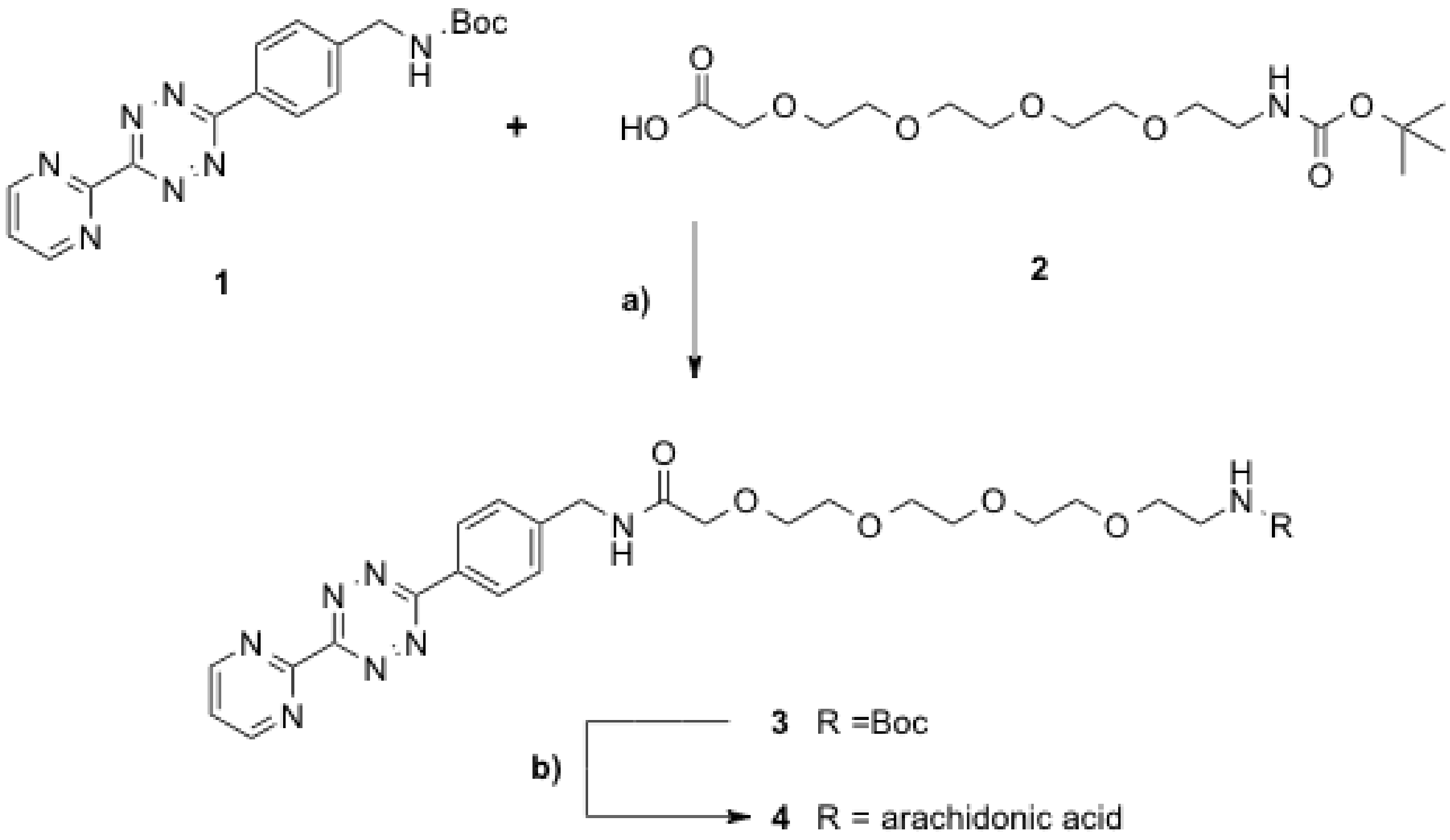

Scheme S1: Synthesis of anandamide-tetrazine **4**. a) 1. TFA, DCM, 0°C; 2. DIPEA, HATU, HOBt, DMF, RT, 42% b) 1. TFA, DCM, 0°C; 2. arachidonic acid, DIPEA, HATU, HOBt, RT, 46%.

*tert*-butyl (4-(6-(pyrimidin-2-yl)-1,2,4,5-tetrazin-3-yl)benzyl)carbamate (**1**) was synthesised according to the procedures described by Willems *et al*.[1]

2,2-dimethyl-4-oxo-3,8,11,14,17-pentaoxa-5-azanonadecan-19-oic acid (**2**) was synthesised like from Shirude *et al* described.[2]

*tert*-butyl (3-oxo-1-(4-(6-(pyrimidin-2-yl)-1,2,4,5-tetrazin-3-yl)phenyl)-5,8,11,14-tetraoxa-2-azahexadecan-16-yl)carbamate (**3**)

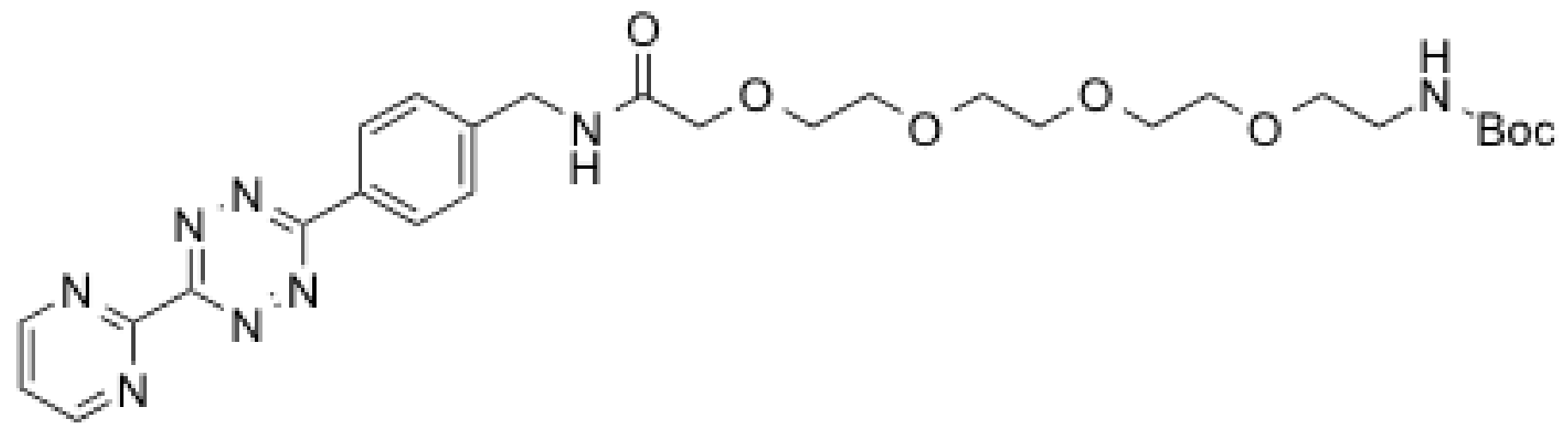

*tert*-butyl (4-(6-(pyrimidin-2-yl)-1,2,4,5-tetrazin-3-yl)benzyl)carbamate (**1**) (162 mg, 0.160 mmol) was dissolved in 6.4 mL DCM and cooled to 0°C before TFA (1.6mL) were added. After 45 min the solvent was removed *in vacuo* and the resulting residue was used in the next reaction without further purification.

2,2-dimethyl-4-oxo-3,8,11,14,17-pentaoxa-5-azanonadecan-19-oic acid (**2**) (228 mg, 0.649 mmol, 2.5 eq.) was diluted in 3.77 mL dry DMF and HATU (119 mg, 0.312 mmol, 1.2 eq.), HOBt (102 mg, 0.780 mmol, 3 eq.) and finally DIPEA (0.192 mL, 0.780 mmol, 3 eq.) were added. After 10 min the deprotected tetrazinamine (0.068 mg, 0.260 mmol, 1 eq.) was added and the reaction was stired over night at RT. The reaction was diluted with DCM and the washed with water and brine before the combined organic phases were dried over $MgSO_4$ and the solvent was removed *in vacuo*. The residue was purified by column chromatography (silica, DCM/EtOAc/MeOH, 5:5:1) to obtain the **3** as violet oil (66.0 mg, 0.110 mmol, 42%).

$\mathbf{R_f}$ = 0.22 ($CH_2Cl_2$/EtOAc/MeOH, 5:5:1).

**$^1$H-NMR** (600 MHz, $CDCl_3$): δ [ppm] = 9.11 (d, $^3J$ = 4.9 Hz, 2H, 2xC-$H_{arom}$), 8.68 (d, $^3J$ = 8.4 Hz, 2H, 2xC2-$H_{arom}$), 7.59 – 7.54 (m, 1H, C-$H_{arom}$), 4.61 (d, $^3J$ = 6.2 Hz, 2H, Ar-C$H_2$-NH), 4.12 – 4.08 (m, 2H, C=O-C$H_2$-O), 3.73 – 3.50 (m, 14H, tetraethylene glycol 7x C$H_2$), 3.47 (t, $^3J$ = 5.1 Hz, 2H, C$H_2$-NH), 1.40 (s, 9H, 3xC$H_3$-*t*Bu).

**$^{13}$C-NMR** (150 MHz, $CDCl_3$): δ [ppm] = 164.57, 163.32, 159.82, 159.81, 158.62, 129.33, 128.77, 122.68, 70.62, 70.36 (Tetraethylenglykol 7 x $CH_2$), 42.76 (Ar-$CH_2$-NH), 28.64 ($CH_3$-tBu).

**HR-MS** (ESI): $[M+Na]^+$ calc.: 621.2755, found: 621.2763.

**FT-IR** (ATR, $cm^{-1}$): 3336 (br, w), 2921 (w), 1702 (m), 1676 (m), 1610 (m), 1563 (m), 1529 (m), 1434 (m), 1380 (vs), 1250 (m), 1144 (m), 1113 (m), 844 (s).

14-((5*Z*,8*Z*,11*Z*,14*Z*)-icosa-5,8,11,14-tetraenamido)-*N*-(4-(6-(pyrimidin-2-yl)-1,2,4,5-tetrazin-3-yl)benzyl)-3,6,9,12-tetraoxatetradecanamide (**4**)

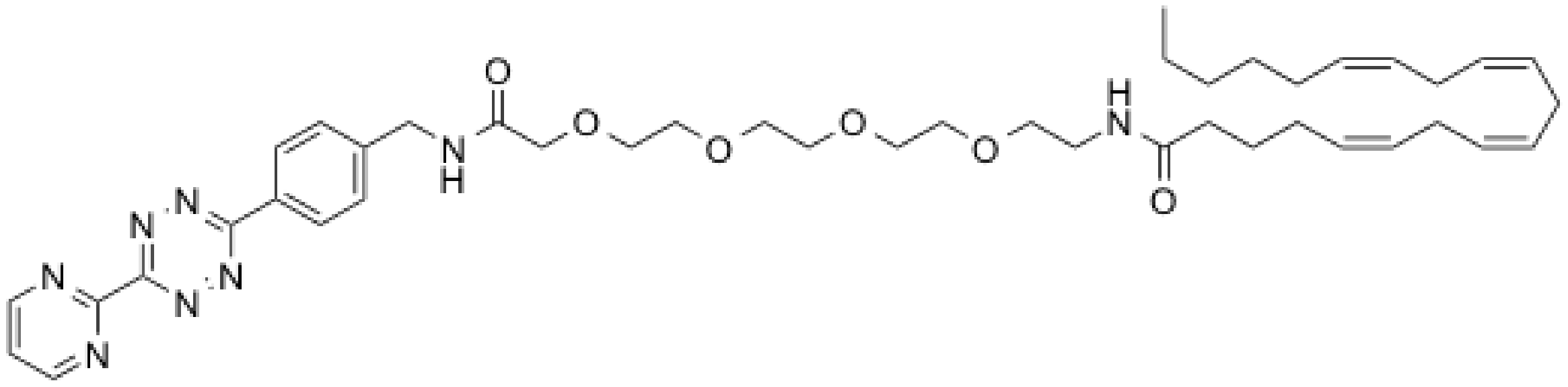

*tert*-Butyl (3-oxo-1-(4-(6-(pyrimidin-2-yl)-1,2,4,5-tetrazin-3-yl)phenyl)-5,8,11,14-tetraoxa-2-azahexadecan-16-yl)carbamat (**3**) (66.0 mg, 0.1103 mmol) was solved in 1.15 mL DCM and 0.3 mL TFA was added. After 45 min the solvent was removed *in vacuo* and the resulting residue was used in the next step without further purification.

Arachidonic acide (54.6 µmol, 166 µmol, 1.5 eq.) was dissolved in 0.7 mL dry DMF and HATU (50.3 mg, 132 µmol, 1.2 eq.), HOBt (44.7 mg, 331 µmol, 3 eq.) and DIPEA (56,3 µL, 332 µmol, 3 eq.) were added. After 10 min the deprotected amine (55.0 mg, 110 µmol, 1 Äquiv.), dissolved in 0.6 mL dry DMF, was added and the reaction was steered for 2h at RT. The reaction was diluted with DCM and washed with saturated $NH_4Cl$-solution. After drying of the combined organic phases over $MgSO_4$ and removing of the solvent *in vacuo*, the residue was purified by column chromatography (silica, *i*Hex/EtOAc, 1:1 → DCM/EtOAc/MeOH, 10:10:1). **4** was received as violet oil (39,6 mg, 50,6 µmol, 46%).

**$R_f$** = 0.72 ($CH_2Cl_2$/EtOAc/MeOH, 2:2:1).

**$^1$H-NMR** (600 MHz, $CDCl_3$): δ [ppm] = 9.11 (d, $^3J$ = 4.8 Hz, 2H, C-$H_{arom}$), 8.67 (d, $^3J$ = 8.4 Hz, 2H, 2xC3'-$H_{arom}$), 7.59 – 7.54 (m, 1H, C5-$H_{arom}$), 5.43 – 5.23 (m, 8H, 8xC*H*), 4.61 (d, $^3J$ = 6.2 Hz, 2H, Ar-C$H_2$-NH), 4.11 – 4.02 (m, 2H, C=O-C$H_2$-O), 3.73 – 3.36 (m, 18H, tetraethylene glykole 8x$CH_2$, C$H_2$-NH), 2.82 – 2.72 (m, 6H, 3xC-$H_{2arach}$), 2.18 – 2.12 (m, 2H, $CH_{2arach}$), 2.09 – 1.99 (m, 4H, 2xC-$H_{2arach}$), 1.67 (q, $^3J$ = 7.5 Hz, 2H, C-$H_{2arach}$), 1.37 – 1.20 (m, 6H, 3xC-$H_{2arach}$), 0.86 (t, J = 7.0 Hz, 3H, C-$H_{3arach}$).

**$^{13}$C-NMR** (100 MHz, $CDCl_3$): δ [ppm] = 173.38 , 170.64, 164.54, 163.27, 159.72 ($C_q$), 158.64 (2x$C_{arom}$), 129.35 (CH), 129.33 (2xC), 128.90, 128.81, 128.74, 128.43, 128.38, 128.06, 127.72 (7xCH), 126.95, 122.75 ($C_{arom}$), 71.30, 70.65, 70.60, 70.47, 70.34, 70.13 (Tetraethylenglykol 8x$CH_2$), 42.73 ($CH_2$-NH), 39.40, 36.21 ($C_{arach}$), 31.72 ($C_{arach}$), 29.92 ($C_{arach}$), 27.43 ($C_{arach}$), 26.91 ($C_{arach}$), 25.85 ($C_{arach}$), 25.83 ($C_{arach}$), 25.74 ($C_{arach}$), 22.79 ($C_{arach}$), 14.30 ($C_{arach}$).

**HR-MS** (ESI): $[M+H]^+$ calc.: 785.4709, found: 785.4727.

**FT-IR** (ATR, $cm^{-1}$): 3311 (m), 2923 (s), 1555 (s), 1413 (s), 1103 (s).

**Click chemistry of norbornene-containing hCA with anandamide tetrazine.** MSNs (MSN-phSA, 0.5 mg) were loaded in 500 µL of calcein solution (1 mM) for 1 h. The loaded particles were collected by centrifugation (14,000 rpm, 16,837 rcf, 4 min) and 500 µL HBSS buffer solution was added. After addition of 0.5 mg norbornene-containing hCA the particles were redispersed and stirred for 1 h. Then, 5 µg anandamide tetrazine (DMSO stock solution, 2 mg/mL) were added and stirred for 1 h respectively. The particles were thoroughly washed with HBSS buffer (4 times), collected by centrifugation (5,000 rpm, 2,200 rcf, 4 min, 15 °C), and finally redispersed in 500 µL HBSS buffered solution.

**Cell Culture.** HeLa cells were grown in Dulbecco's modified Eagle's medium (DMEM):F12 (1:1) (lifeTechnologies) with Glutamax I medium and KB cells in folic acid deficient Roswell Park Memorial Institute 1640 medium (RPMI 1640, lifeTechnologies), both supplemented with 10 % fetal bovine serum (FBS) at 37 °C in a 5 % $CO_2$ humidified atmosphere. The cells were seeded on collagen A-coated LabTek chambered cover glass (Nunc). For live cell imaging the cells were seeded 24 or 48 h before measuring, at a cell density of 2x104 or 1x104 cells/$cm^2$.

The FGF-2 and EGF dependant neural stem cell line ENC1 was derived from E14 mouse embryonic stem cells and cultured as described.[9] ENC1 cells were maintained in gelatine coated flasks and propagated in a 1:1 mixture of Knockout-DMEM (*Life Technologies*) and Ham's F-12 (*Sigma*) supplemented with 2 mM GlutaMAX-I (*Life Technologies*), 100 U/ml penicillin (*Sigma*), 100 mg/ml streptomycin (*Sigma*) 1% N2 and 20 ng/ml each of mouse recombinant FGF-2 and EGF (*Peprotech*). N2 supplement was produced in house as described, with the exception that Insulin was of human origin (*Sigma* I9278) instead of bovine.[10]. Stem cells were seeded on ibidi 8-well µ-slides.

***In vitro* Cargo Release.** Cells were incubated 7 – 24 h prior to the measurements at 37 °C under a 5% $CO_2$ humidified atmosphere. Shortly before imaging, the medium was replaced by $CO_2$-independent medium (Invitrogen). During the measurements all cells were kept on a heated microscope stage at 37 °C. The subsequent imaging was performed as described in the spinning disk confocal microscopy section.

**Endosomal compartment staining.** To stain the early/late endosome or the lysosome with GFP, commercially available CellLight© staining from lifeTechnologies was used . The cells were simultaneously incubated with MSNs and the BacMam 2.0 reagent. The concentration of the labeling reagent was chosen with 25 particles per cell (PCP) of the BacMam 2.0 reagent (cf. staining protocol [4]). For incubation, the cells stayed at 37 °C under 5% $CO_2$ humidified atmosphere for 21 – 24 h till the measurement.

**Caspase-3/7 staining.** For apoptosis detection commercially available CellEvent™ Caspase-3/7 Green Detection Reagent was used. A final concentration of 2.5 µM Caspase-3/7 reagent and 0.5 µg/mL Hoechst 33342 were added to the cells for 30 min and imaging was performed without further washing steps.

**Uptake studies.** The functionality of the folic acid targeting ligand was evaluated in a receptor competition experiment. For this purpose, one part of the KB cells was pre-incubated with 3 mM folic acid, to block the receptors, for 2 h at 37 °C under a 5% $CO_2$ humidified atmosphere. Then the KB cells were incubated with particles for 2/5/8 h at 37 °C under a 5% $CO_2$ humidified atmosphere. For staining the cell membrane, the cells were incubated with a final concentration of 10 µg/mL wheat germ agglutinin Alexa Fluor 488 conjugate for one minute. The cells were washed once with $CO_2$-independent medium and imaged. For stem cell uptake studies cells were seeded the day prior to incubation. They were incubated for 2 h with free anandamide-tetrazine at a final concentration of 10 µg/ml. After 2 h 15 µg of particles were added and incubated for another 2 h. Then, the cells were washed 3x with medium containing growth factors and if preincubated free anandamide-tetrazine and incubated until imaging. Immediately before imaging, cells membranes were stained using cell mask deep red (lifetechnologies) and washed with medium.

**Spinning disc confocal microscopy.** Confocal microscopy for live-cell imaging was performed on a setup based on the Zeiss Cell Observer SD utilizing a Yokogawa spinning disk unit CSU-X1. The system was equipped with a 1.40 NA 100x Plan apochromat oil immersion objective or a 0.45 NA 10x air objective from Zeiss. For all experiments the exposure time was 0.1 s and z-stacks were recorded. DAPI and Hoechst 33342 were imaged with approximately 0,16 $W/mm^2$ of 405 nm, GFP was and the caspase-3/7 reagent were imaged with approximately 0.48 $W/mm^2$ of 488 nm excitation light. Atto 633 was excited with 11 $mW/mm^2$ of 639 nm. In the excitation path a quad-edge dichroic beamsplitter (FF410/504/582/669-Di01-25x36, Semrock) was used. For two color detection of GFP/caspase-3/7 reagent or DAPI/Hoechst 33342 and Atto 633, a dichroic mirror (560 nm, Semrock) and band-pass filters 525/50 and 690/60 (both Semrock) were used in the detection path. Separate images for each fluorescence channel were acquired using two separate electron multiplier charge coupled device (EMCCD) cameras (PhotometricsEvolveTM)

## Further Characterization

MSNs containing thiol-functionality exclusively at the external particle surface (sample MSN-SH) were established following a previously described delayed co-condensation approach.[5] In a second step, benzene sulfonamide (phSA) groups were covalently attached to the silica nanoparticles via a short bifunctional crosslinker (maleimide-$C_6$-NHS) at mild reaction conditions (sample MSN-phSA). After cargo loading, the enzyme CA was added to the buffered particle solution (pH 7.4). The formation of the inhibitor-enzyme complex (phSA-CA) leads to a dense coating at the external particle surface (MSN-phSA-CA).

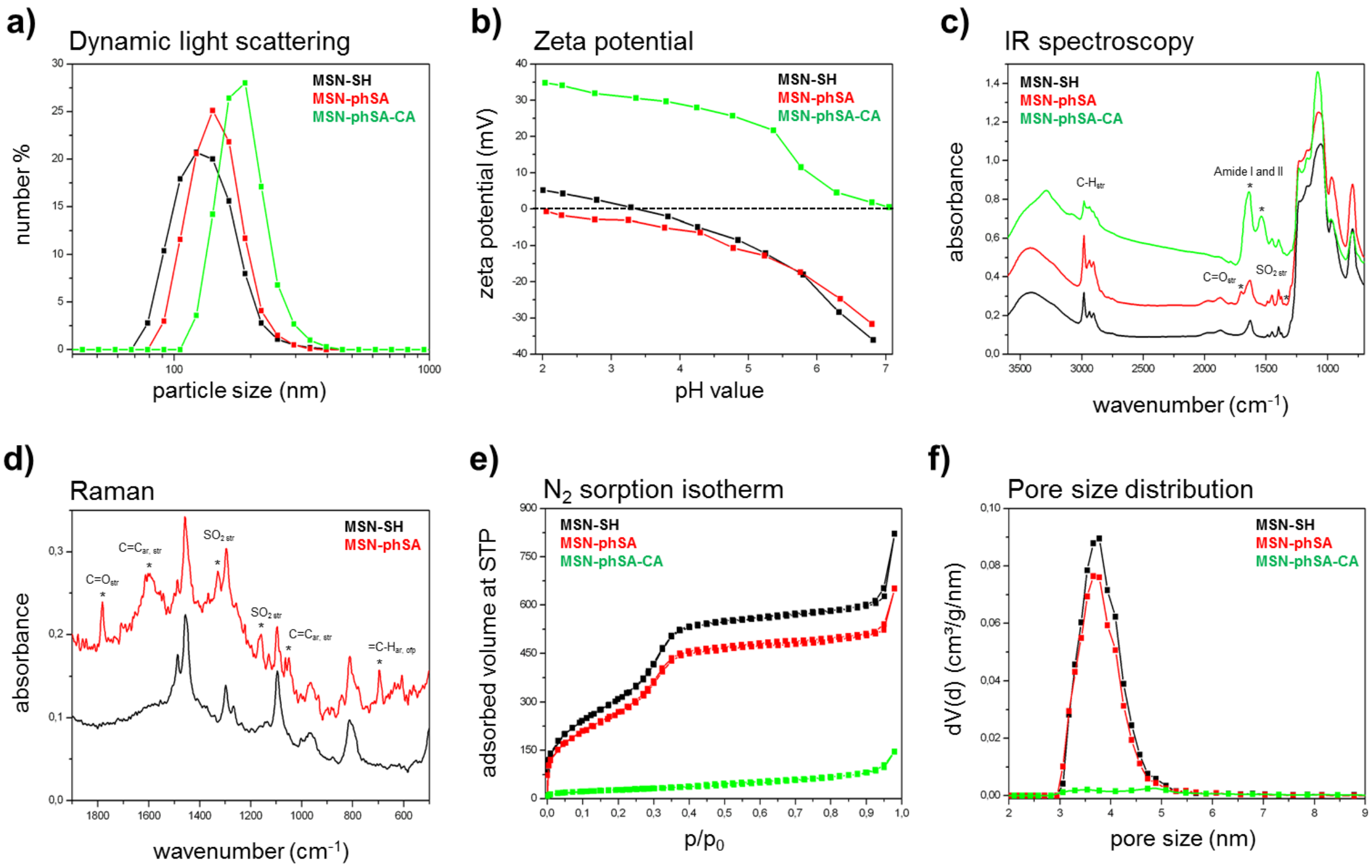

**Figure S2: Characterization of multifunctional MSNs. a) Dynamic light scattering (DLS), b) zeta potential measurements, c) infrared (IR) spectroscopy data, d) Raman spectroscopy data, e) nitrogen sorption isotherms, and f) DFT pore size distribution of the MSNs. MSN-SH (black), MSN-phSA (red) and MSN-phSA-CA (green).**

Dynamic light scattering (DLS) measurements showed the size distribution of the functionalized MSNs to be narrow and around 150 nm (Figure S2a), implying excellent colloidal stability after all functionalization steps. The surface charge of silica nanoparticles, measured as the zeta potential, changed due to the stepwise attachment of organic moieties (Figure S2b): The isoelectric point (IEP) of MSN-SH (pH 3.6) was shifted to a more acidic pH value (< 2) for MSNs containing the benzene sulfonamide groups on the outer surface. The tendency for sulfonamide groups to be protonated is relatively low due to the stabilizing resonance effect, which leads to the increase in negative surface charge (predominantly influenced by silanol content). After attachment of the carbonic anhydrase, a drastic increase of the zeta potential was observed resulting from amino acid residues that can be easily protonated - such as arginine, histidine and lysine - on the surface. IR data for all samples showed typical vibrational modes of the silica framework between 780 and 1300 $cm^{-1}$ (Figure S2c). MSNs containing the benzene sulfonamide groups showed additional modes for C=O stretching vibrations at 1700 and 1627 $cm^{-1}$ and a peak of weak intensity at 1340 $cm^{-1}$, which belongs to the typical asymmetric $SO_2$ stretching vibration modes of the sulfonamide groups. For the sample MSN-phSA-CA, amide vibrations (Amide I: 1639 $cm^{-1}$, C=O stretching vibration; Amide II: 1535 $cm^{-1}$, N-H deformation and C-N stretching vibration) of high intensity were observed; these are typical for proteins. Raman spectroscopy provided data complementary to IR spectroscopy. In Figure S2d a more detailed view of the spectra for MSN-SH and MSN-phSA in the range between 1900 and 600 $cm^{-1}$ is depicted and various additional bands (*) were observed for the benzene sulfonamide-functionalized particles. (data for MSN-phSA-CA not shown, for full range Raman spectra see Figure S5). Nitrogen sorption measurements show type IV isotherms for MSN-SH and MSN-phSA, confirming mesoporosity of the silica nanoparticles. Relatively high surface areas (up to 1200 $m^2/g$) and pore volumes (0.8 $cm^3/g$) were observed for MSN-SH and MSN-phSA (Table S4).

**Table S4: Porosity parameters of functionalized MSNs.**

| Sample | BET surface area ($m^2/g$) | Pore volume[a] ($cm^3/g$) | DFT pore size[b] (nm) |
|---|---|---|---|
| MSN-SH | 1170 | 0.83 | 3.8 |
| MSN-phSA | 1004 | 0.72 | 3.7 |
| MSN-phSA-CA | 99 | 0.07 | - |

[a]Pore volume is calculated up to a pore size of 8 nm to remove the contribution of interparticle porosity. [b]DFT pore size refers to the peak maximum of the pore size distribution.

Importantly, the DFT pore size distribution (Figure S2f) was not affected by the attachment of the benzene sulfonamide linkers and no incorporation of organic groups inside the mesopores was observed. The attachment of the bulky enzyme carbonic anhydrase resulted in a drastic reduction of surface area and pore volume for sample MSN-phSA-CA. Thus, the carbonic anhydrase enzymes were able to efficiently block the mesopores even towards the access of nitrogen molecules. We observed no pore size distribution for MSN-phSA-CA in the range between 2 and 9 nm. This confirms the successful synthesis of carbonic anhydrase-coated MSNs via benzene sulfonamide linkers.

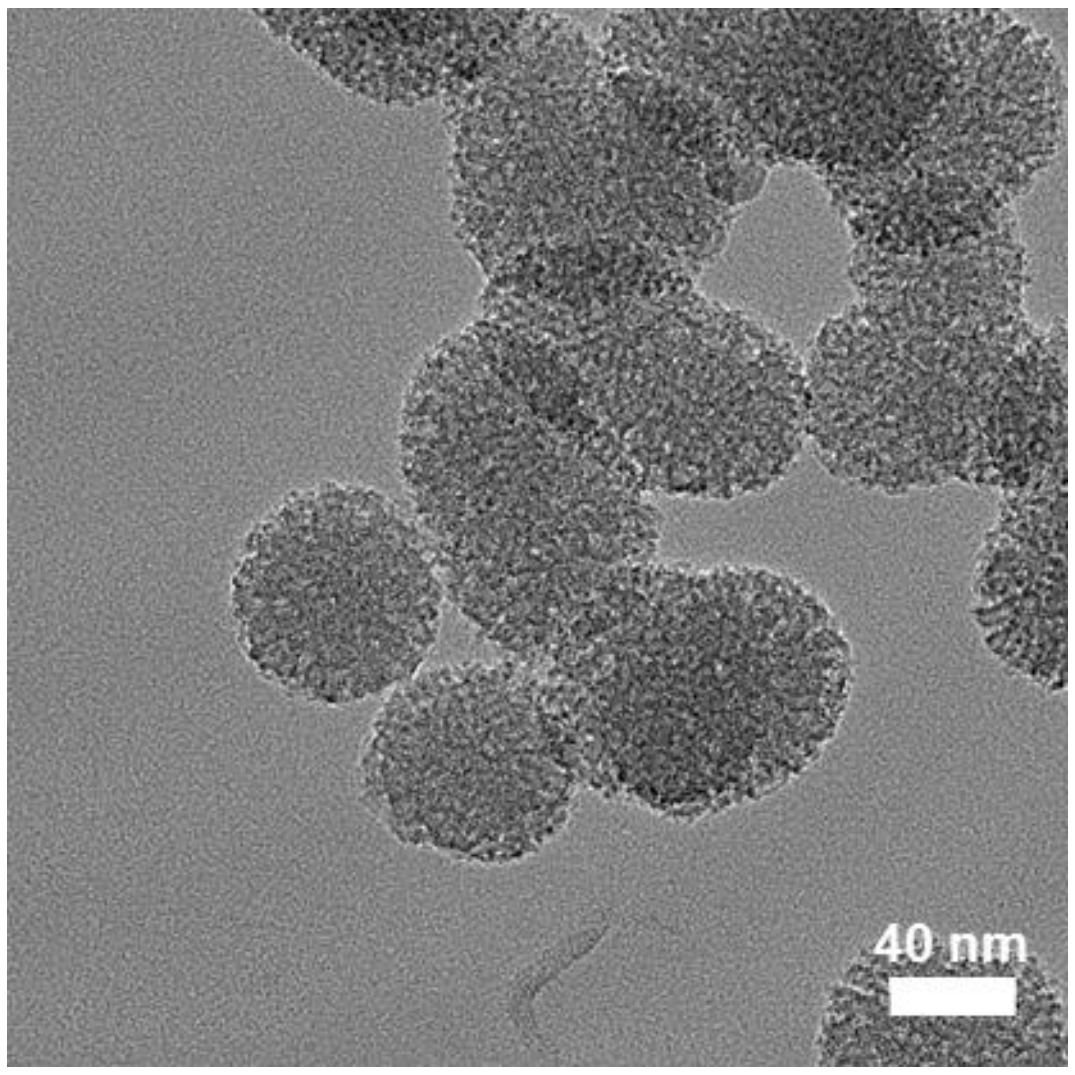

**Figure S3: Transmission electron micrograph of thiol-functionalized MSNs (MSN-SH).**

TEM images of thiol-functionalized MSNs are depicted in Figure S3 and display mostly spherically shaped particles with a radially disposed worm-like structure of the mesopores. The mesoporous structure is also confirmed by the first-order reflection of the mesoporous material observed with small-angle X-ray diffraction (XRD) (Figure S4).

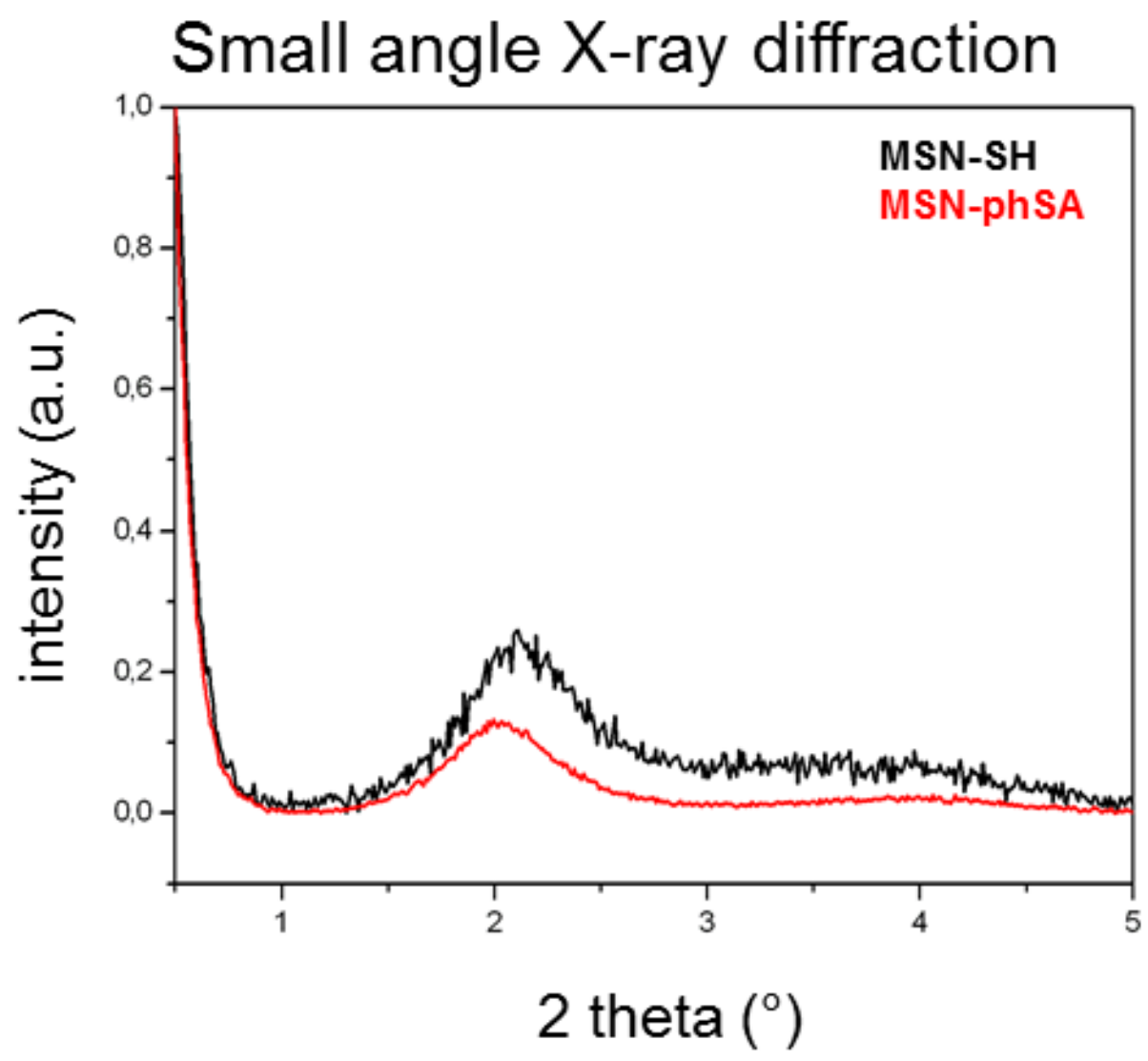

**Figure S4: Small-angle X-ray diffraction pattern of MSN-SH (black) and MSN-phSA (red).**

Full-range Raman spectra are shown in Figure S5. The sharp band at 1780 $cm^{-1}$ can be assigned to the carbonyl stretching vibration of the diacylamine group of the maleimide residue that is covalently attached to the thiol groups of the MSN surface. The presence of phenyl groups is confirmed by characteristic bands of aromatic C=C stretching vibrations (1600 $cm^{-1}$ and 1055 $cm^{-1}$) and aromatic =C-H out-of-plane deformation vibrations (693 $cm^{-1}$). The broadening of the signal at 1600 $cm^{-1}$ can be assigned to a partial overlap by amide II vibration modes. The bands at 1326 and 1156 $cm^{-1}$ are related to the characteristic asymmetric and symmetric stretching vibrations of the sulfonamide group, respectively.

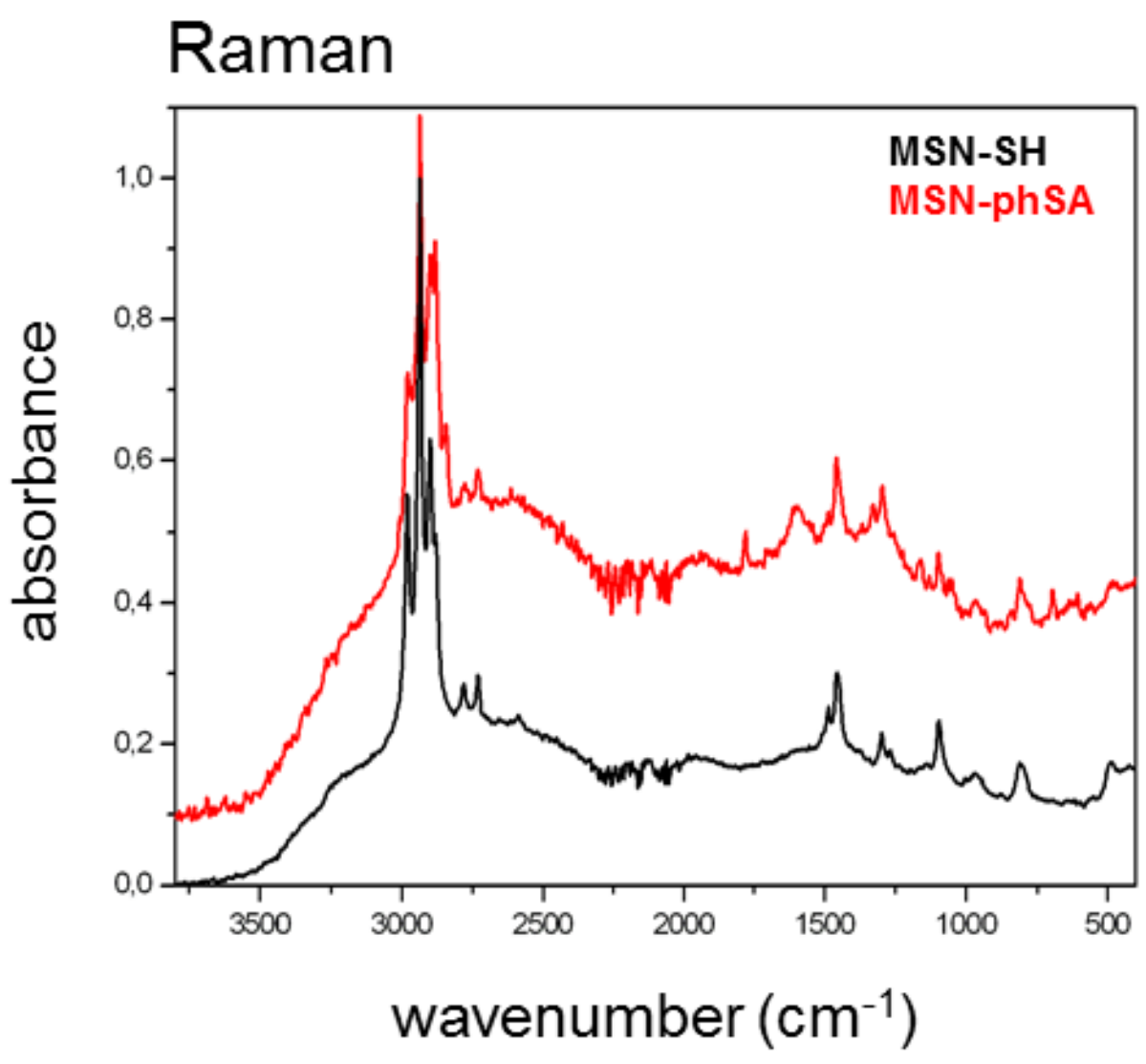

**Figure S5: Raman spectroscopy data of functionalized MSNs. MSN-SH (black) and MSN-phSA (red). For clarity reasons, the spectra are shifted along the y-axis by 0.1 units. Measurements were performed at an excitation wavelength of 633 nm.**

The amount of attached organic moieties on the MSNs was investigated by thermogravimetric analysis, showing an additional weight loss of about 14 % after attachment of CA (TGA, Figure S6). MSN-phSA particles show an additional weight loss of 3 % in the range between 130 and 900 °C due to the attachment of the benzene sulfonamide linker and enzyme-coated MSNs (MSN-phSA-CA) feature a relatively high additional weight loss compared to sample MSN-SH (+14 % at 900 °C). Apparently, degradation of the carbonized enzymes occurs only at very high temperatures, and is not even finished at 900 °C. This was already observed before for thermogravimetric analysis of enzyme-coated MSNs.

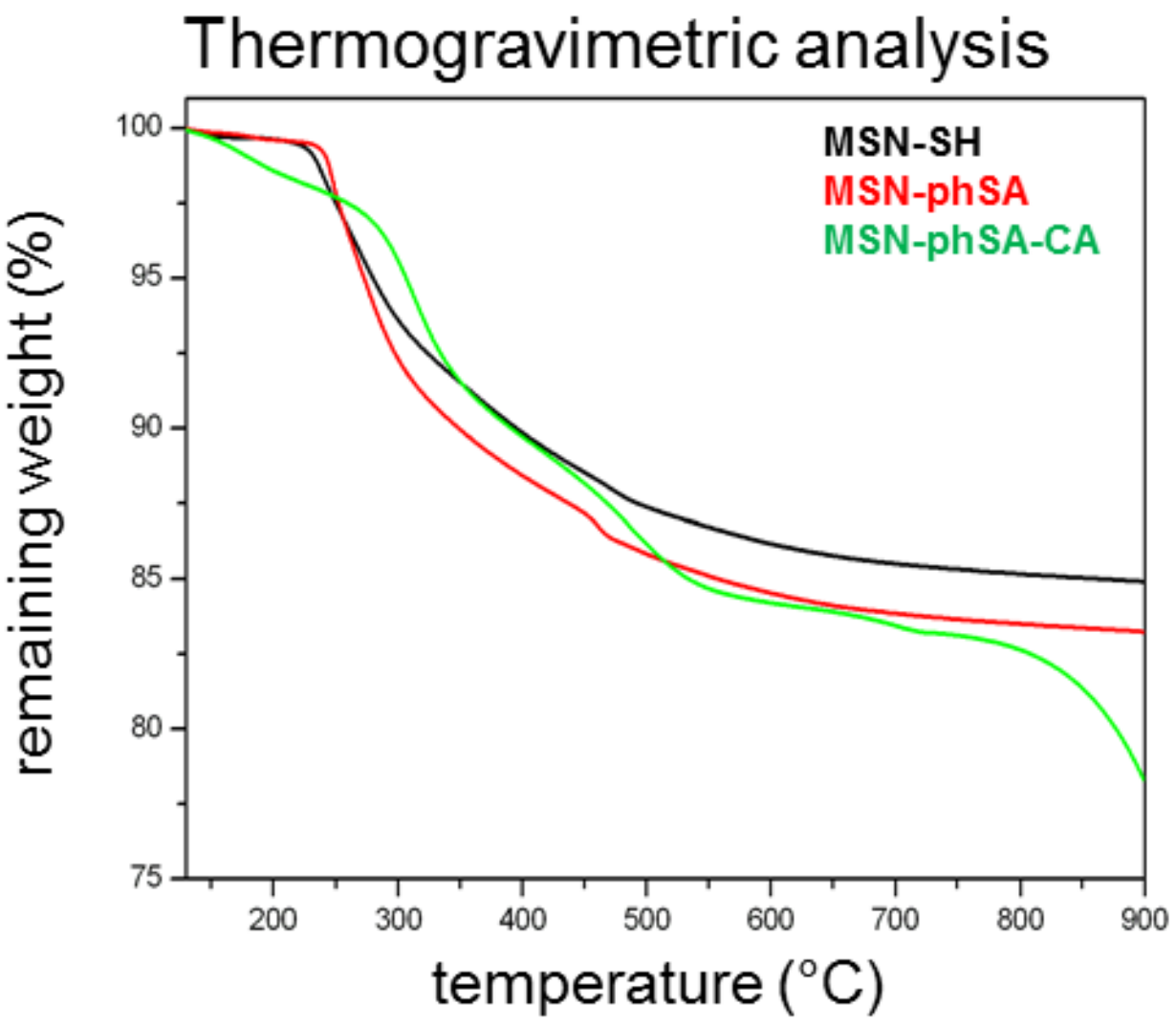

**Figure S6: Thermogravimetric analysis of MSN-SH (black), MSN-phSA (red), and MSN-phSA-CA (green).**

In order to investigate the pH-responsive removal of the bulky gatekeepers from the particles, *in vial* cargo release experiments were performed. We used a custom-made two-compartment system to analyze the time-based release of the fluorescent model cargo fluorescein.[6] After incorporation of fluorescein molecules into the mesoporous system, carbonic anhydrase was added to block the pore entrances. An efficient sealing of the pores and no premature release of the cargo was observed for the sample MSN-phSA-CA dispersed in HBSS buffer (pH 7.4) at 37 °C (Figure S7a, closed state, black curve). After 3 h the solution was exchanged and the particles were dispersed in citric-acid phosphate buffer (CAP buffer, pH 5.5). The change to acidic milieu, which simulates the acidification of endosomes, causes a significant increase in fluorescence intensity over time (open state, red curve). Furthermore, we could show the long-term stability of the capping system for more than 16 hours in HBSS buffer and cell medium at pH 7.4 (Figure S7b). These *in vial* release experiments demonstrate efficient sealing of the pores with carbonic anhydrase acting as a bulky gatekeeper, preventing premature cargo release and allowing for release upon acid-induced detachment of the capping system.

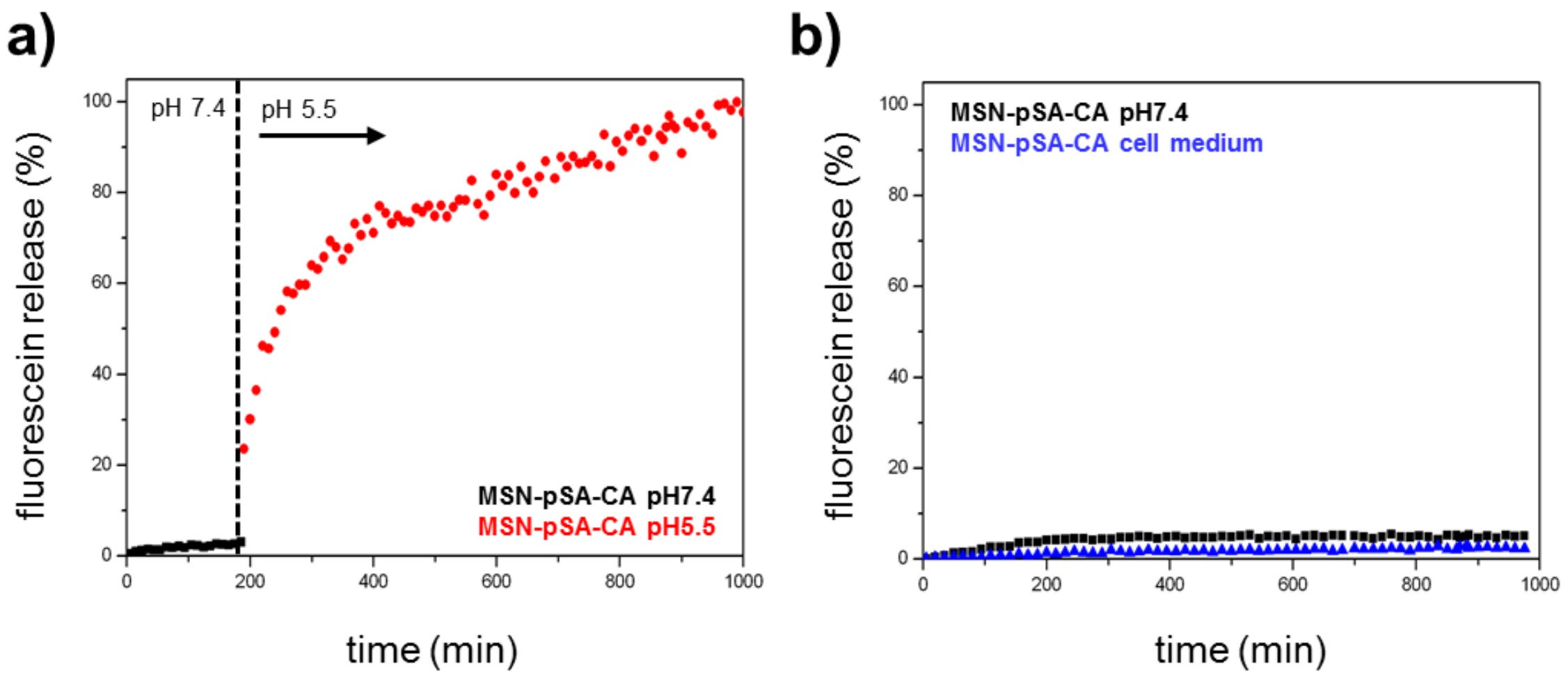

**Figure S7:** ***In vial*** **release kinetics of fluorescein molecules from the enzyme-coated MSNs at different pH values. (a) Sample MSN-pSA-CA features no premature release of the fluorescent cargo molecules in HBSS buffer solution at pH 7.4 (closed state, black curve). After 3 h the medium was changed to slightly acidic milieu (CAP buffer, pH 5.5, red curve) resulting in a significant increase in fluorescence intensity. The gatekeepers are detached from the particle surface upon acidification, causing an efficient and precisely controllable release of fluorescein from the mesoporous system. (b) Long-term stability of the capping system was investigated in HBSS buffer (pH 7.4, black curve) and cell medium (blue curve). No unintended cargo release was observed within about 16 h.**

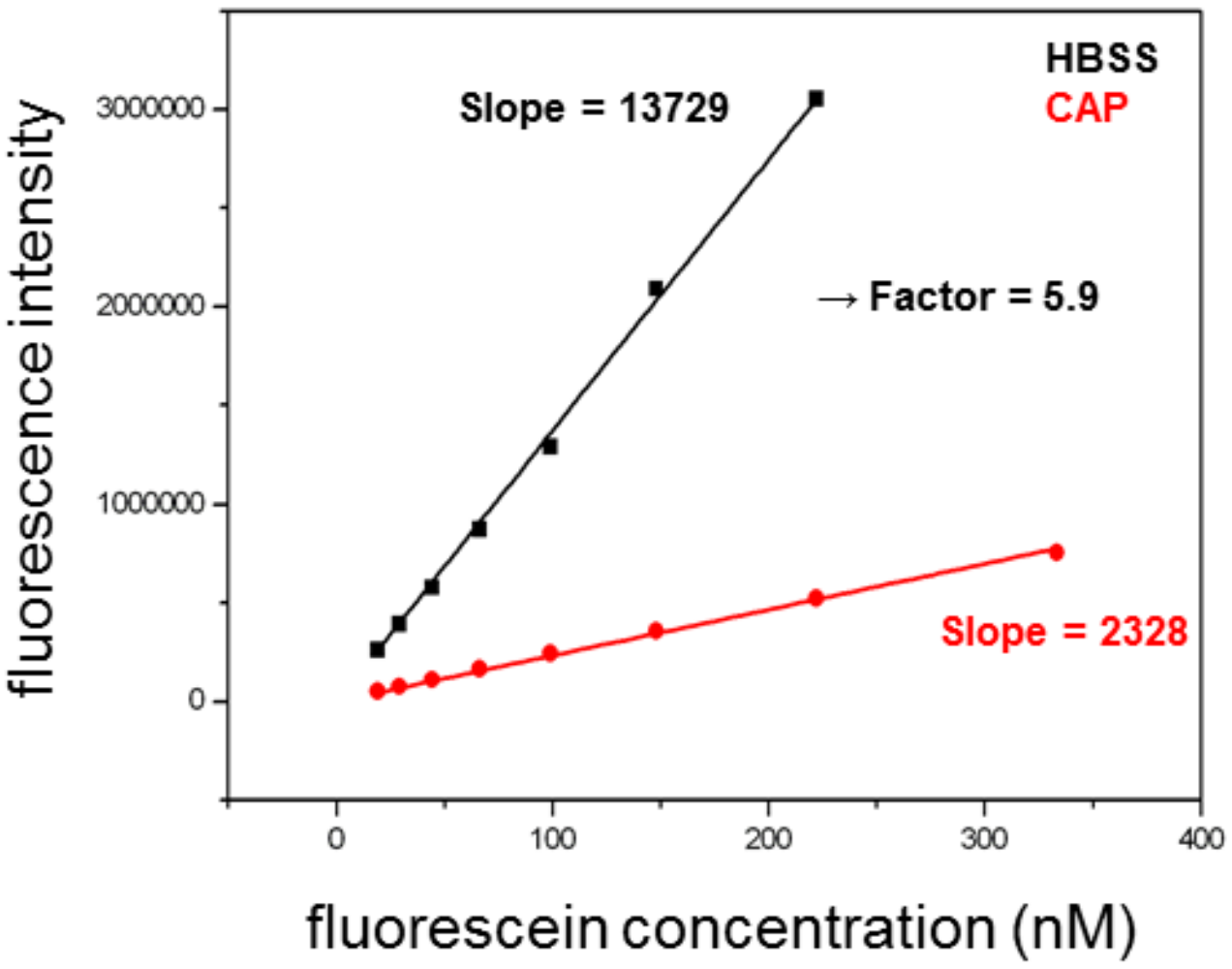

**Figure S8: Calibration curves of fluorescein at pH 7.4 (HBSS buffer, black curve) and pH 5.5 (CAP buffer, red curve).**

The enzyme activity assay investigates the hydrolysis in TRIS-buffered solution of a chromogenic substrate (p-nitrophenyl acetate, NPA) in the presence of the enzyme, generating nitrophenol. UV-Vis spectroscopy is used to measure the resulting absorption maximum at 400 nm. Figure S9 shows the resulting curve for the non-catalyzed (no carbonic anhydrase) reaction, which can be taken as baseline. The slight slope for this curve is due to the hydrolysis rate of the pure substrate in aqueous solution in the absence of catalytic enzymes. In the presence of 100 nM enzyme (non-inhibited) the maximum conversion of the substrate can be obtained. A slight decrease in conversion efficiency can be observed upon addition of 50 µg of MSN-SH particles due to marginal reduction of enzyme activity in the presence of silica nanoparticles. We assume that this effect corresponds to minor unspecific attachment of the carbonic anhydrase to the silica nanoparticles causing blocking of the active sites to some extent. In comparison, the addition of inhibitor-containing particles (MSN-phSA) causes a significant decrease of the slope of the resulting curve. This proves a specific formation of the inhibitor-enzyme complex at the external surface of the silica nanoparticles. Thus we have shown conclusively that the sulfonamide-functionalized MSNs are able to bind the enzyme carbonic anhydrase. At neutral pH values, the enzyme is specifically attached to the sulfonamide-functionalized particle surface resulting in an inhibition of the enzyme's active site. This leads to a drastic decrease in enzyme activity.

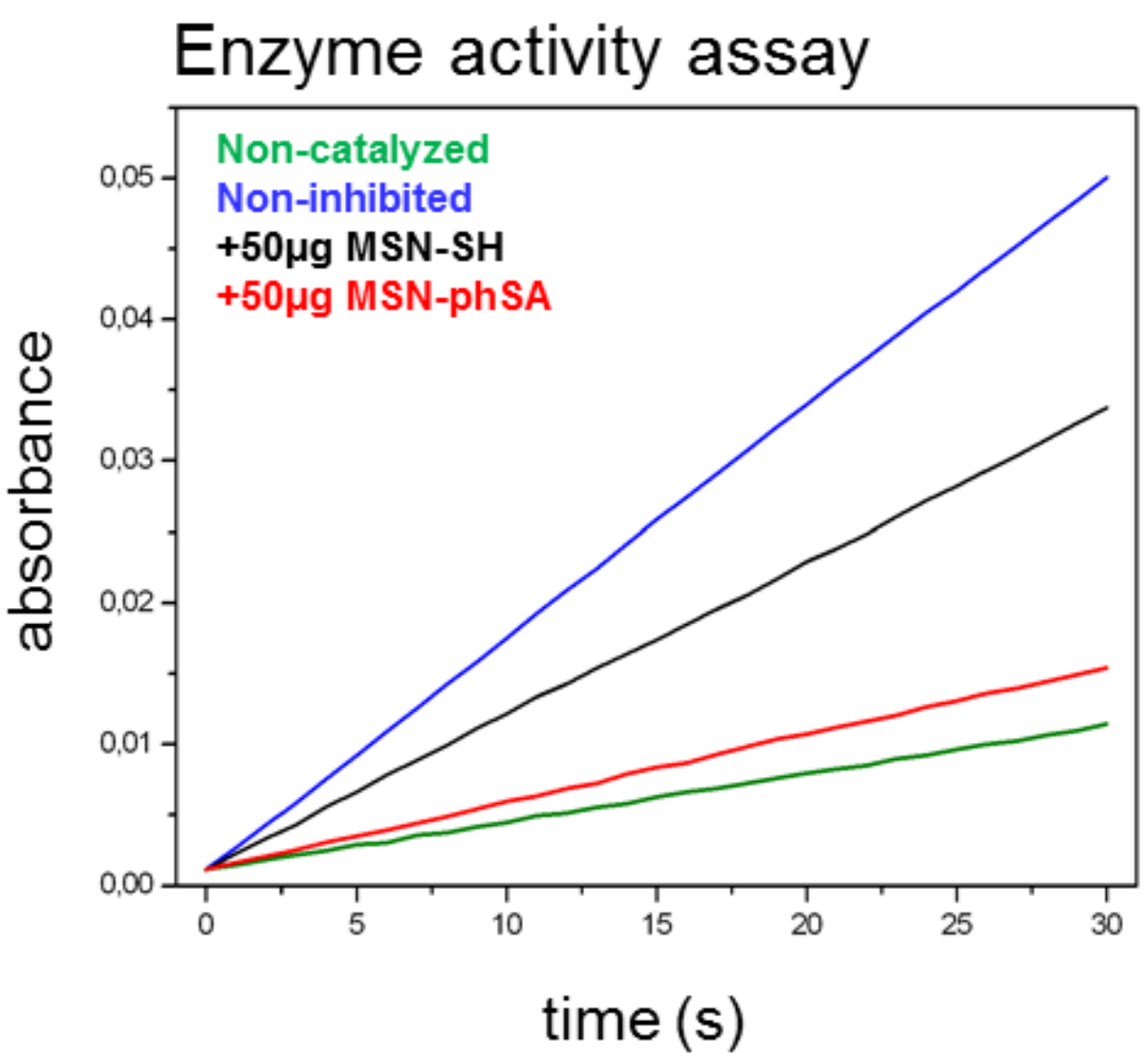

**Figure S9: Enzyme activity assay of CA catalyzing the hydrolysis of the chromogenic substrate p-nitrophenyl acetate measured by UV-Vis spectroscopy (absorbance at 400 nm). Non-catalyzed (green) and non-inhibited (blue) reaction and after addition of MSN-SH (black) or MSN-phSA (red).**

Employing fluorescent live-cell imaging, we investigated the *in vitro* release behavior of encapsulated 4',6-diamidino-2-phenylindole (DAPI) in HeLa cancer cells. The molecular size of DAPI is similar to fluorescein. It was therefore expected to efficiently enter the mesoporous system of the silica nanoparticle. Due to its effective *turn-on* fluorescence upon intercalation into DNA double strands, DAPI is commonly used as nuclei counterstain in cell imaging (about 20 fold enhancement in fluorescence intensity).[7] Since DAPI is cell membrane permeable, free fluorescent dye molecules are able to stain the nucleus within very short time periods (1-5 min), as described in several staining protocols.[8] After incorporation of DAPI into the mesoporous system of the silica nanocarriers, the pores were sealed by addition of carbonic anhydrase. The HeLa cells were incubated for a total time period of 24 h with the loaded particles, which were additionally labeled with Atto 633 dye (red), as depicted in Figure S10. After 7 h of incubation, MSNs were efficiently taken up by the cells and were found to be located in endosomes.

Importantly, almost no staining of the nuclei with DAPI (blue) could be observed at this time point. Only after 15 h, blue fluorescence (even more intensive after 24 h) provided evidence of efficiently released DAPI from the MSNs. Control experiments in which the sample supernatant after particle separation (centrifugation) was added to the HeLa cells showed no significant nuclei staining even after 24 h (Figure S10d). These cell experiments prove a substantial time-dependent release of DAPI from the mesopores of our nanocarrier system and also show that no free dye molecules were present in the solution. We suggest that the observed delayed nuclei staining results from a cascaded release mechanism. First, acidification throughout the endosomal pathway to late endosomes or endolysosomes is of key importance. Only the pH change to mildly acidic values (about 5.5) makes the detachment of the bulky gatekeepers from the MSN hosts possible. Subsequent opening of the pores leads to an efficient cargo release.

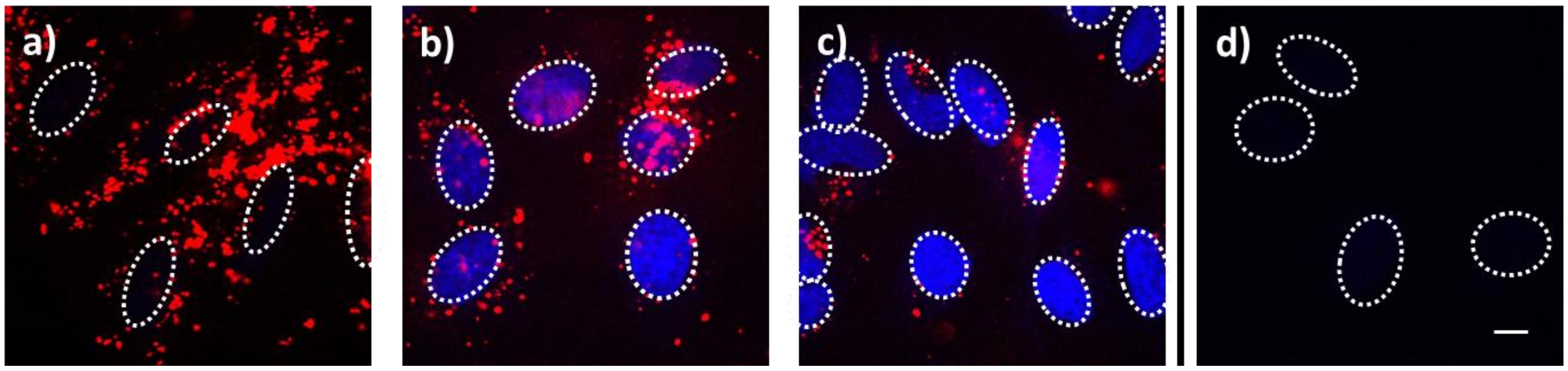

**Figure S10: Fluorescence microscopy of HeLa cells incubated with MSN-phSA-CA nanoparticles loaded with DAPI and labeled with Atto 633 (red) after a) 7 h, b) 12 h and c) 24 h of incubation. The delayed nuclei staining with DAPI (blue) is caused by a time-dependent release of DAPI based on the need for an acidic environment. d) In a control experiment, the incubation with the supernatant solution (without MSNs) showed no staining of the nuclei with DAPI after 24 h, suggesting that no free DAPI molecules were present in the particle solution. The nuclei are indicated with dashed circles. The scale bar represents 10 µm.**

In order to verify the fate of our drug delivery vehicles ending up in acidic cell compartments, co-localization experiments with labeled MSNs and endosomes or lysosomes were performed. Simultaneous with particle incubation, the HeLa cells were transfected with a BacMam reagent in order to express different fusion-constructs of green fluorescent protein (GFP) and early/late endosome or lysosome markers, respectively. After 24 h of incubation with fluorescently labeled nanoparticles, almost no co-localization (yellow) between early endosomes and MSNs could be observed, as can be seen in Figure S11a. In contrast, multiple yellow spots indicating co-localization between GFP (green) and MSNs (red) were clearly visible in the case of late endosomal and lysosomal staining after 21 h (Figure S11b/c, denoted by arrows). This shows that the localization of our nanocarriers in an acidic compartment is crucial to initiate cargo release.

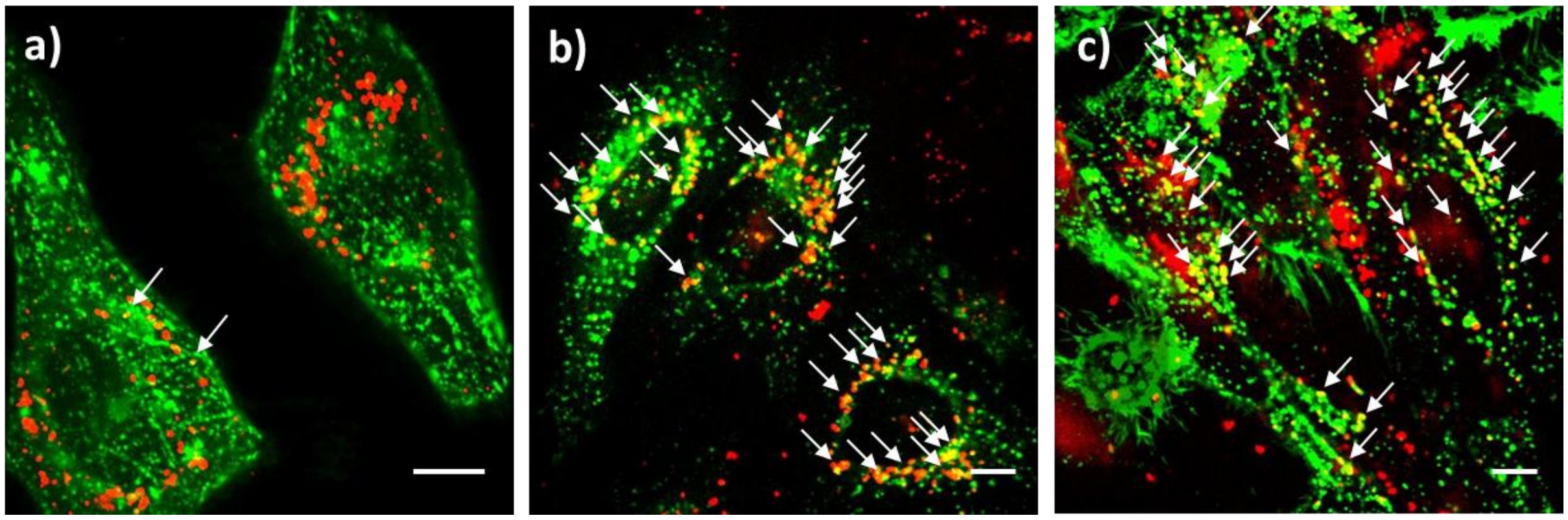

**Figure S11: Fluorescence microscopy of HeLa cells incubated with Atto 633-labeled MSN-phSA-CA (red) after a) 24 h on GFP-early endosome (green) tagged cells, b) 21 h incubation on GFP-late endosome (green) tagged cells, and c) 21 h on GFP-lysosome (green) tagged cells. Co-localization (yellow) could be primarily observed for late endosomes and lysosomes (indicated with arrows) suggesting that the multifunctional MSNs are located in acidic compartments after endocytosis. The scale bar represents 10 µm.**

## Stem Cell targeting

To test targeting of anandamide functionalized particles neural stem cells were treated with anandamide functionalized particles. As control, cells were pretreated with free anandamide-tetrazine and incubated with anandamide particles after 2 h of pretreatment. Another control was performed with control particles without anandamide functionalization. After two hours of particle incubation all cells were washed with medium. Already a few hours after incubation anandamide-particles were observed to stick to the cells in large amounts whereas particles without anandamide did not show this behavior. After 24h anandamide-targeted particles were successfully taken up into the cells. Cells that were incubated with control particles or preincubated with anandamide did not show as much particle uptake (Fig. S12).

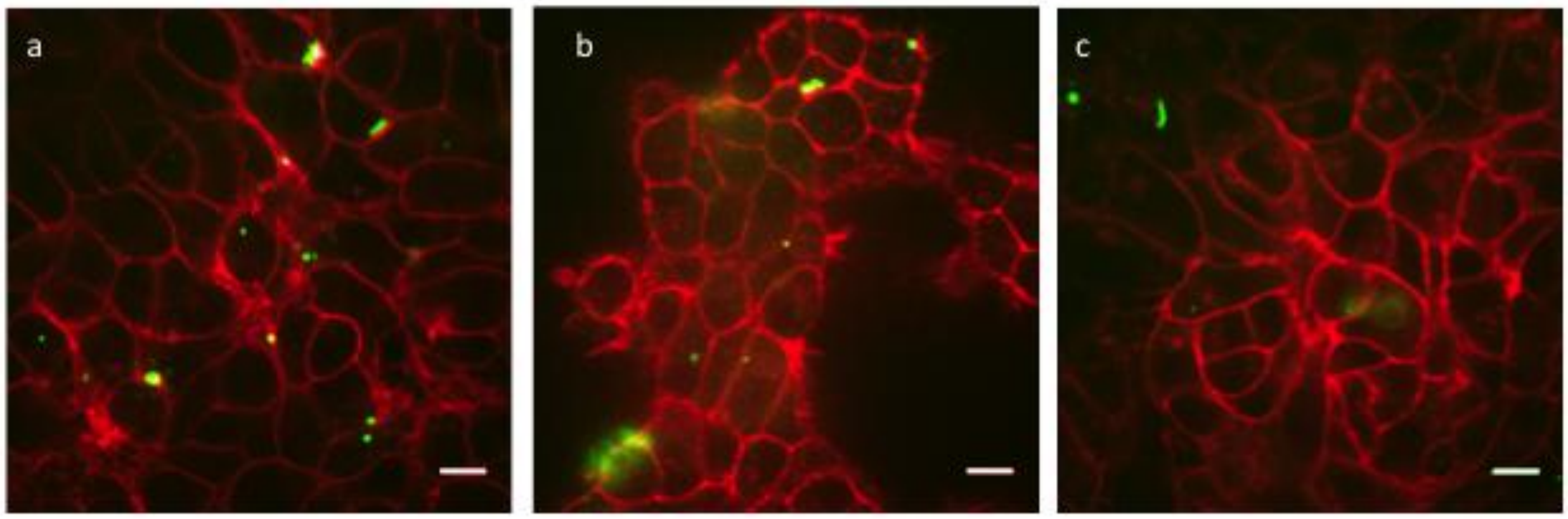

**Figure S12: Fluorescence microscopy of neural stem cells incubated with calcein-labeled MSN-phSA-CA (green) after 24h a) incubated with anandamide-targeted MSN-phSA-CA, b) pretreated with free anandamide-tetrazine and incubated with anandamide targeted MSN-phSA-CA afterwards c) incubated with control MSN-phSA-CA without anandamide. Cell membranes are stained with cell mask deep red. The scale bar represents 10 µm.**